\documentclass[fleqn,usenatbib]{mnras}
\usepackage{newtxtext,newtxmath}

\usepackage[T1]{fontenc}

\usepackage{graphicx,amsmath}
\usepackage{caption,subcaption,bm}
\usepackage[dvipsnames]{xcolor}

\newcommand{\Sec}[1]{Sect.~\ref{sec:#1}}
\newcommand{\Section}[1]{Section~\ref{sec:#1}}

\newcommand{\Fig}[1]{Fig.~\ref{fig:#1}}
\newcommand{\Figure}[1]{Figure~\ref{fig:#1}}

\newcommand{\Tab}[1]{Tab.~\ref{tab:#1}}
\newcommand{\Table}[1]{Table~\ref{tab:#1}}

\newcommand{\Eqn}[1]{Eqn.~(\ref{eq:#1})}

\newcommand{\beq}{\begin{equation}}
\newcommand{\eeq}{\end{equation}}

\newcommand{\etaO}{\eta_{\rm O}}
\newcommand{\etaH}{\eta_{\rm H}}
\newcommand{\etaA}{\eta_{\rm A}}

\newcommand{\Mdotw}{\dot{M}_{\rm w}}
\newcommand{\Mdota}{\dot{M}_{\rm a}}
\newcommand{\Mdotv}{\dot{M}_{\rm v}}

\newcommand\ergpersec{\,\rm erg\,s^{-1}}

\newcommand\yr{\,\rm yr}
\newcommand\Myr{\,\rm Myr}

\newcommand\au{\,\rm au}

\newcommand\Msun{\,M_\odot}
\newcommand\Msunyr{\,M_\odot\,{\rm yr}^{-1}}

\newcommand\ergstb{\!\!\big[{\rm erg/s}\big]\!\!}
\newcommand\Msunyrtb{\!\!\big[10^{-8}\!\Msun/{\rm yr}\big]\!\!}
\newcommand\chk{\checkmark}
\newcommand\nop{$\times$}

\newcommand\mns{\hspace{-5pt}-}

\newcommand\betap{\beta_{\rm P}}
\newcommand\logb{\log{\betap}}
\newcommand\loglx{\log{L_X}}
\newcommand\Feff{F_{\rm eff}}
\newcommand\fgfm{|F_{\rm gas}/F_{\rm mag}|}

\newcommand{\ee}[1]{\times 10^{#1}}

\usepackage{color,soul,ulem}

\definecolor{dark-red}{rgb}{0.75, 0.00, 0.00}
\definecolor{hlcolor}{rgb}{1.00, 0.94, 0.92}\sethlcolor{hlcolor}

\renewcommand\emph[1]{\textit{#1}}

\usepackage{xspace}
\newcommand{\nirv}{\textsc{nirvana-iii}\xspace}

\newcommand{\V}{\mathbf{v}}
\newcommand{\B}{\mathbf{B}}
\newcommand{\E}{\mathbf{\mathcal{E}}}
\newcommand{\LX}{L_{\rm X}}

\newcommand{\vA}{v_{\rm \!A}}
\newcommand{\vp}{v_{\rm p}}
\newcommand{\Bp}{B_{\rm p}}

\newcommand{\reyn}{\rm Reyn}
\newcommand{\maxw}{\rm Maxw\vphantom{y}}

    \title[Hall-MHD of X-ray photoevaporative disc winds]%
          {Hall-magnetohydrodynamic simulations of X-ray photoevaporative
           protoplanetary disc winds}

    \author[E.~Sarafidou~et al.]{%
        Eleftheria~Sarafidou,$^{1,2}$\thanks{E-mail: <elsarafidou@aip.de> (AIP)}
        Oliver~Gressel,$^{1,3}$
        Giovanni~Picogna,$^{4}$
        and Barbara~Ercolano$^{4}$\vspace{4pt}\\
        $^{1}$Leibniz-Institut f\"ur Astrophysik Potsdam (AIP),
              An der Sternwarte 16, 14482, Potsdam, Germany\\
        $^{2}$Institut f\"ur Physik und Astronomie, Universit\"at Potsdam,
              Karl-Liebknecht-Str. 24/25, 14476 Golm, Germany\\
        $^{3}$Niels Bohr International Academy, The Niels Bohr Institute,
              Blegdamsvej 17, DK-2100, Copenhagen \O, Denmark\\
        $^{4}$Universit\"ats-Sternwarte,
              Ludwig-Maximilians-Universit\"at M\"unchen,
              Scheinerstr. 1, 81679 M\"unchen, Germany
    }

    \date{Accepted XXX. Received YYY; in original form ZZZ}

    \pubyear{2015}

\begin{document}

\label{firstpage}
\pagerange{\pageref{firstpage}--\pageref{lastpage}}
\maketitle


\begin{abstract}
  Understanding the complex evolution of protoplanetary disks (PPDs) and their dispersal via energetic stellar radiation are prominent challenges in astrophysics. It has recently been established that specifically the X-ray luminosity from the central protostar can significantly heat the surface of the disk, causing powerful photoevaporative winds that eject a considerable fraction of the disc's mass. Recent work in the field has moreover shown the importance of global PPD simulations that simultaneously take into account non-ideal magnetohydrodynamic (MHD) effects and detailed thermochemistry.
  Our motivation with the current paper lies in combining these two aspects and figure out how they interact. Focus is put on the Hall Effect (HE) and the impact it has on the overall field topology and mass loss/accretion rates. Utilizing a novel X-ray temperature parametrisation, we perform 2D-axisymmetric MHD simulations with the \nirv fluid code, covering all non-ideal effects. We find that, in the aligned orientation, the HE causes prominent inward displacement of the poloidal field lines that can increase the accretion rate through a laminar Maxwell stress. We find that outflows are mainly driven by photoevaporation -- unless the magnetic field strength is considerable (i.e., $\beta_p\leq 10^{3}$) or the X-ray luminosity low enough (i.e., $\loglx\leq 29.3$). Inferred mass loss rate are in the range of the expected values $10^{-8}$ to $10^{-7}\Msunyr$. For comparison, we have also performed pure hydrodynamic (HD) runs and compared them with the equivalent MHD runs. Here we have found that the magnetic field does indeed contribute to the mass loss rate, albeit only discernibly so for low enough $L_X$ (i.e., $\loglx\leq 30.8$). For values higher than that, the wind mass loss predicted from the MHD set converges to the ones predicted from pure HD.
\end{abstract}

\begin{keywords}
  accretion, accretion discs -- magnetohydrodynamics -- protoplanetary discs, circumstellar matter -- stars: pre-main-sequence -- X-rays: stars
\end{keywords}


\section{Introduction} \label{sec:intro}

Protoplanetary discs (PPDs) are rotating discs of gas and dust surrounding young pre-main sequence stars that are formed from the gravitational collapse of a molecular cloud. These discs have been observed to have a wide range of sizes, masses, and lifetimes. Observations of young clusters find that the fraction of stars surrounded by PPDs declines from $85\%$ at ages $\leq 1\,\Myr$ down to $\leq 10\%$ at $5-10\,\Myr$ \citep{Ribas2014} -- which sets the rough time frame for the formation of planetary systems. The mass of the discs can be estimated to be between $10^{-3}$ and $10^{-1} \Msun$ \citep{andrews2005circumstellar,2023ASPC..534..539M}. Taken together, this suggests that a process is in place which allows for a very rapid clearing of the disc, once its dispersal has been established. Moreover, timescales for gas-giant planet formation have been found to be comparable to the clearing of the disc  \citep{helled2013giant}, suggesting that the processes of disc dispersal via mass loss and that via gap-opening are intertwined \citep[see][for a comprehensive review]{ercolano2017dispersal}. Understanding how the mass of the disc is accreted\,/\,lost will thus greatly contribute to our general understanding of disc evolution and, ultimately, planet formation.

The main mechanism that can account for the disc's mass loss is that related to so-called photoevaporative winds (PEWs). These are thermally driven winds that cause the shedding of the disc, effectively acting as mass sink. PEWs are driven by the heating effects of high-energy radiation within the disc gas. When the heated surface layer becomes gravitationally unbound, it is launched away in a disc wind \citep[see][for a recent review discussing both theoretical and observational aspects of disc outflows]{2023ASPC..534..567P}. In a typical PPD, there are believed to be three main contributing radiation types: extreme-UV (EUV), far-UV (FUV), and X-rays emitted from the central star \citep{hollenbach2000disk}.
Arguably, EUV seems to take a secondary role in the launching of the wind, providing quite a small wind mass loss rate of about $\Mdotw\simeq10^{-10}\Msunyr$ \citep{alexander2006photoevaporation}. At the same time, models with FUV suggest that the wind mass loss is originating in the outer regions of the disc \citep{ercolano2017dispersal}. The models presented here reach out to $\simeq 20\au$, and we chose to focus on photoevaporation due to X-ray heating. During this process, X-ray are absorbed by the material in the disc, leading to the ionisation of gas molecules and the creation of free electrons and ions. The disc material become thermally unbound and is then launched away from the disc and into the surrounding interstellar medium. \citep{alexander2013dispersal}. Typically, a wind mass loss rate of $\Mdotw\sim 10^{-8}\Msunyr$ is expected \citep{owen2010radiation}. However, since the driving force of the wind is always pointing in the radial direction, the transfer of angular momentum from the disc to the outflowing material is purely advective in PEWs.

Mass accretion to the central star is achieved through angular momentum transport. For some time the prevailing theory was that this is a result of turbulence in the disc. Hydrodynamic instabilities, albeit present, cannot account for the observed accretion rates \citep{armitage2011dynamics}. Another promising candidate for driving turbulence, the magneto-rotational instability \citep[MRI; ][]{balbus1991powerful} was proven to be substantially suppressed in weakly ionised PPDs \citep[see, e.g.,][and references therein]{fleming2003local,gressel2015global}. Searching for a more efficient mechanism that could drive angular momentum transport, the focus has recently been shifting back to the old idea of magnetic/magnetocentrifugal winds \citep[][]{2000prpl.conf..759K} that remain viable in the realm of non-ideal MHD.

Magnetic winds can broadly be classified into two types, namely magnetic pressure driven winds and magneto-centrifugal winds (MCWs), as discussed in detail in Appendix C of \citet{bai2017global}. The former are generated by the vertical magnetic pressure gradient exerted by the winding-up toroidal field \citep[e.g.,][]{lynden1996magnetic}, while the latter result from the poloidal magnetic field being anchored to the disc \citep[e.g.,][]{blandford1982hydromagnetic}. In-between these two limiting cases, there exists a continuum of configurations that can be parameterised by their so-called magnetic lever arm, a measure for the ability of the magnetic field configuration to exert a torque on the outflowing material.
MCWs can be a means for an efficient vertical transport of angular momentum, and therefore they are currently considered essential to explain mass accretion.

When considering the role of magnetic fields in PPD evolution, it appears mandatory that non ideal MHD effects should be taken into account. Most regions of PPDs are too cold for sufficient thermal ionisation, and effective ionisation may be achieved only in the discs' surface layer \citep{gammie1996layered,stone2000transport}. The ionisation fraction of PPDs is expected to remain low from the midplane to the outer regions of the disc, where ionisation sources (radioactivity and cosmic rays) are either relatively weak or shielded by the overlying material. As a consequence, this situation leads to very low ionisation fractions $\chi = n_e/n \leq 10^{-10}$ \citep{perez2011surface}.

The non ideal effects acting in PPDs are ~i) Ohmic Resistivity (OR), ~ii) Ambipolar Diffusion (AD) and ~iii) the Hall Effect (HE). The relative importance of the three effects can be roughly delineated by their dependence on the specific field strength, $B/\rho$. This means that OR is mainly operating in high density areas, that is, in the midplane and the inner disc. It has been shown that in these regions, OR critically affects the turbulence otherwise caused by the MRI \citep{fleming2003local} and is largely responsible for the so-called ``dead zone'' (in terms of magnetic activity) region \citep{gammie1996layered}. Adding AD in the computational models results to the additional suppression of the MRI in the upper and outer layers of the disc \citep{bai2013windI,gressel2015global} that were previously thought to be ``active''. These findings further strengthen the idea that angular momentum transport via magnetic turbulence is far weaker that initially thought. At the same time, even when including both OR and AD, the upper (well ionised) layers retain viable conditions for launching a magnetic wind from any suitable weak poloidal magnetic field \citep{bai2013windII}, thus providing a robust alternative for angular momentum transport in laminar discs threaded by weak magnetic fields. While OR and AD are both diffusive phenomena\footnote{The latter can act in a non-diffusive way in combination with Keplerian shear, though, see \citet{2004ApJ...608..509D} for the respective MRI dispersion relation.}, the same is not true for the Hall Effect, which is non-dissipative and leads to a reorientation of the field direction.

In the context of PPDs, this property leads to a curious bifurcation. In concert with the rather dominant shearing-out of radial fields caused by the differential Keplerian rotation, the HE makes the gas dynamics dependent on the relative orientation of the magnetic field \citep[see, e.g.,][studying the impact of this bifurcation on linear MRI]{wardle2012hall}. In the case where the magnetic field is aligned\footnote{Note that this convention is tied to the case of $\etaH>0$, which is typically the dominant sign in the largely dust-free ionisation model considered here.} with the rotational axis of the disc, the radial components of the field are in general amplified (\textit{aligned HE}). In the opposite case, the radial components are generally reduced (\textit{anti-aligned HE}), and the MRI can, for instance, be squelched. This duality of the HE has an effect in magnetic winds as well: it has been shown that in the case of aligned HE, we get increased radial transport of angular momentum \citep{bai2014hall,lesur2014thanatology}. This can be explained as a consequence of the radially inward ``pinching'' of the vertical field due to the Hall shear instability \citep[HSI; ][]{2005astro.ph..7313R,2008MNRAS.385.1494K} in that case.

Global models of discs that include at least OR and AD are relatively scarce -- and even less are the ones that combine them with photoevaporative effects. Initially, \citet{gressel2015global} performed global axisymmetric simulations with OR and AD, and with a fixed temperature structure. They found out that AD is suppressing the MRI and that the accretion rates observed are a result of a magnetic wind. Next \citet{bai2017hall} and \citet{bethune2017global} performed global simulations with the addition of the Hall effect (albeit with different assumptions in their ionisation models). They established that wind launching is indeed a consistent outcome of these simulations. \citet{bai2017global} included a tabulated approach for the microphysics and confirmed that when the magnetic field is aligned with the rotational axis of the disc, the HE enhances the radial component of the field resulting to higher accretion rates via magnetic winds. Subsequent refinements by \citet{wang2019global} and  \citet{gressel2020global} implemented self consistent thermochemistry and the latter even considered heating from ambipolar collisions. The thermochemistry of their models was mainly based on processes driven by FUV radiation from the central star. Likewise, \citet{rodenkirch2020global} have performed similar global MHD models for a range of magnetic field strengths. In contrast to the two previously mentioned papers, they chose not to include FUV radiation, but rather focus on stellar EUVs and X-rays. Moreover, they considered OR as the only non ideal MHD effect. The accretion rates observed in \citet{wang2019global} are quite similar to the ones predicted in \citet{bai2017global}; additionally they showed that $\Mdotw$ is strongly affected by the high-energy photon luminosities. \citet{gressel2020global} observed similar behaviour in their model, although in contrast with \citet{bai2017global} and \citet{wang2019global} they observe magnetocentrifugal winds instead of winds driven by magnetic pressure gradient \citep[see  appendix C of][]{bai2017global}.

In the present work, we choose to include all three non ideal MHD effects and include X-ray photoevaporation from the central star. The goal is to take a closer look at the dual nature of the Hall effect in that situation, and how it affects the general gas dynamics of the disc. This paper is structured as follows: in Sections~\ref{sec:methods} and \ref{sec:diag}, we initially present our numerical model and present the key diagnostics we chose, respectively. Next, in \Section{results}, we focus on the dual nature of the Hall Effect (see \Sec{he_cases}) and how that affects the disc dynamics and the outflow. Continuing we investigate how the wind nature and its characteristics are altered by changing the magnetic field strength of the disc and the X-ray luminosity from the central star (see \Sec{param}). Finally, in \Section{concl}, we present our conclusions and summarise our findings.

\section{Methods} \label{sec:methods}

We here perform MHD simulations of the inner regions of PPDs, that are a derivative of the non-ideal MHD disk setup presented in \citet{gressel2020global}. The focus of our current investigation is to take into account X-ray thermodynamic heating (as opposed to FUV-driven) simultaneously with non-ideal magnetohydrodynamic aspects -- notably including the HE. To this end, we have combined our Hall-MHD framework \citep[see][]{2018ApJ...865..105K} with the self-consistent temperature and ionisation structure parametrisation presented in \citet{picogna2019dispersal}.

Our simulations are $2D$ axisymmetric with a spherical-polar coordinate system $(r, \theta, \phi)$ - radius, co-latitude and azimuth respectively. They were performed with the single-fluid \nirv code, which is built around a standard  second order--accurate finite volume--scheme \citep{ziegler2004adi, ziegler2016chemical}. We are using a modified version of the \nirv code that was presented in \citet{gressel2020global}. For updating the grid cells, we use the Harten-Lax-van~Leer Riemann Solver \citep{harten1983upstream} with the Hall term added as described in \citet{Lesur2014}. We are primarily interested in the inner regions of the PPD where the disc is considered to be laminar under typical conditions. Our computational domain covers an area of $r\in(0.75, 22.5)\au$, $\theta\in(0, \frac{1}{2}\pi)$ with a standard grid resolution of $N_r\times N_{\theta} = 576\times 256$.

\subsection{Equations of Motion}

We solve the equations of motion for mass-, momentum- and total energy-density in conservation form along with the induction equation for the magnetic field, that is,
\begin{eqnarray}
  \partial_t\rho+\nabla\!\cdot(\rho\V) & = & 0\,,
  \nonumber\\
  \partial_t(\rho\V)+\nabla\!\cdot\big[\,\rho\V\V
    +P \mathbf{I}-\B\B\,\big] & = & \!-\rho\nabla\Phi\,,
  \nonumber\\
  \partial_t e + \nabla\!\cdot
     \big[\, (e+P)\V - (\V\cdot\B)\B\, \big]
     & = & \!-\rho(\nabla\Phi)\!\cdot\V + \nabla\!\cdot\!\mathcal{S} \,,
     \nonumber\\
  \partial_t \B - \nabla\times(\V\times\B) & = & - \nabla\times\E\,.
     \nonumber
\end{eqnarray}
where the total pressure, $P$, is defined as the sum of the gas and the magnetic pressure, and $\mathcal{S} \equiv \E\times\B$ is the Poynting flux, with $\E$ representing the electromotive force due to the three non-ideal MHD effects, that is
\beq
  \E \equiv \etaO\mathbf{J} \,+\, \etaH\mathbf{J}\times\hat{\mathbf{B}} \,+\, \etaA\mathbf{J}\times\hat{\mathbf{B}}\times\hat{\mathbf{B}}.
\eeq
with  $\etaO, \etaA$ and $\etaH$ the diffusion coefficients of Ohmic Resistivity, Ambipolar Diffusion and Hall Effect, respectively:
\begin{subequations}\label{eq:diffusivities}
  \begin{gather}
    \etaO = \frac{c^2\gamma_e m_e \rho}{4\pi\rho^2 n_e}\,,\qquad
    \etaA = \frac{B^2}{4\pi\gamma_i \rho\rho_i}\,,\qquad
    \etaH  = \frac{cB}{4\pi e n_e} \tag{\theequation a-c}
  \end{gather}
\end{subequations}
where $n_e$, $e$, and $m_e$ are the electron number density, charge and mass, respectively, and $\rho$ is the density of neutrals, $\rho_i$ the ion density, $\gamma = \langle\sigma u\rangle/(m_n+m_i)$, the drift coefficient between ion and neutrals. Moreover, $\langle\sigma u\rangle$ is the ion--neutral collision rate, $\sigma$ the conductivity, which in the case of a proton-electron plasma is $\sigma = n_e e^2/(m_e f_c)$ with $f_c$ the collision frequency, and finally $\mathbf{J} = \nabla\times\B$ is the electric current density \citep[see, e.g.,][]{1999MNRAS.303..239W}.


\subsection{Disc Model \& Boundary Conditions}\label{sec:disk_model}

The equilibrium structure \citep[see][]{2013MNRAS.435.2610N} of our models is described by a power law relationship of the locally isothermal temperature $T$, with the cylindrical radius, $R$:
\begin{equation}\label{eq: eqTemp}
  T(R) =  T_0\left(\frac{R}{R_0}\right)^q\,,
\end{equation}
and with a similar relation of the midplane density with $R$, that is,
\begin{equation}\label{eq: eqrho_mid}
  \rho_{\rm mid}(R) =  \rho_0\left(\frac{R}{R_0}\right)^p\,,
\end{equation}
where we have chosen a temperature slope of $q= -0.5$ (resulting in a mildly flared disc height as a function of radius), and a density slope of $p = -2.25$ -- akin to typical observed PPD profiles \citep[e.g.,][]{2023ASPC..534..539M}. The solutions of the disc equilibrium are:
\begin{eqnarray}
  \rho(\mathbf{r}) & = & \rho_{\rm mid}(R) \exp{\left[\frac{GM_\star}{c_s^2} \left( \frac{1}{r}-\frac{1}{R}\right)\right]}\,,
  \label{eq: dens_eq_sol}\\
  \Omega(\mathbf{r}) & = & \Omega_{\rm K}(R)\, \big[\,(p+q)\left(\frac{H}{R}\right)^2+(1+q)-\frac{qR}{r}\,\big]^{1/2}\,,
  \label{eq: Omega_sol}
\end{eqnarray}
where $\Omega_{\rm K}(R) \equiv (GM_\star R^{-3})^{1/2}$ is the Keplerian angular velocity, and $c_s^2(R) = c_{s0}^2(R/R_0)^q$ is the sound speed squared, and $H(R_0) = c_{s0}/\Omega_K(R_0) = 0.055\,R_0$ fixes the overall pressure scale height of the initial gas disc. Unless otherwise indicated, the initial density profile was scaled to $\Sigma_0 = 200\,{\rm g\,cm^{-2}}$ at $R_0 = 1\au$, and the aspect ratio is $H(R) = 0.055(R/R_0)^{1/4}$, the mass of the central star $M_\star = 0.7M_{\odot}$, the X-ray luminosity of the star is $\LX = 2\times10^{30}\,{\rm erg\,s^{-1}}$. The adiabatic index was chosen as $\gamma = 5/3$, corresponding to an ideal atomic gas. We furthermore assume a mean molecular weight of $\mu \approx 1.37$, representing a typical abundance of the light elements.

\smallskip We use the exact same boundary conditions as \citet{gressel2020global}, which were described in detail in section~3.2.1 of that paper. In brief, we apply an ``outflow'' criterion for the momentum density, and extrapolate the mass density with the power-law slope of the equilibrium model at radial boundaries. Standard damping buffer zones are implemented at the inner and out domain boundary. For the parallel magnetic field components, we generally enforce vanishing gradients and reconstruct the perpendicular component from the divergence constraint. As an exception to this, we set the azimuthal components of both the field and current to zero at the outer radial boundary. The axial boundary is determined by the demanded axisymmetry of our simulations, and the latitudinal boundary in the midplane simply applies mirror symmetry to each variable.


\subsection{X-Ray Heating}\label{sec:xray_heating}

Apart from the inclusion of the HE, the current work is most distinctly different from \citet{gressel2020global} in that it replaces the FUV thermochemistry (plus radiative transfer) with a heating-prescription due to X-ray irradiation from the central star -- with the purpose of studying the transition from purely (X-ray) photoevaporative to progressively more magnetic disc winds.

Our approach follows the line of work pioneered by \citet{owen2010radiation} and recently refined by \citet{picogna2019dispersal}. The parametrisation assumes that in regions that are permeated by stellar X-rays, the gas temperature, $T_{\rm gas} = T_{\rm X}(\Sigma,\xi)$, is solely given as a function of the intervening column density, $\Sigma$, and the ionisation parameter $\xi$. The justification of this approach is explained in detail in \citet{ercolano2022modelling} and the references quoted therein.

In brief, the ionisation parameter is introduced as $\xi\equiv \LX/(n\,r^2)$ where $n$ is the local number density of the gas and $\LX$ is the assumed X-ray luminosity. We employ a look-up table approach for the function $T_{\rm X}(\Sigma,\xi)$, which is derived from complex thermal\,/\,ionisation calculations performed with the \textsc{mocassin} code \citep{ercolano2008x} and stored in tabulated form \citep[using the same coefficients as][]{picogna2019dispersal}.

The heating term is implemented as a source term in the energy equation, following a simple ''$\beta$~cooling'' (aka Newtonian relaxation) approach where the local gas temperature is relaxed towards the target temperature $T_{\rm gas}\rightarrow T_{\rm X}$ on a timescale that is a small fraction of the local orbital time. We moreover adopt a specific treatment for the transition into the disc interior, which is essentially shielded from exterior radiation. In the region where the X-rays become progressively less dominant, we expect the assumption of near-instantaneous relaxation to break down. We thus define a transition for a value of $\log_{10}(\xi)=-7$, and smoothly taper both the target temperature and the relaxation timescale for the $\beta$~cooling in that region. We, moreover, implement two possibilities for the disc interior, ~i) a comparatively slow $\beta$~cooling \citep[similar to][]{gressel2015global}, or ~ii) a fully adiabatic scenario, where the source term to the energy equation wanes completely. Unless otherwise noted, we use the former as the default. Finally, we have confirmed that our implementation in \nirv broadly reproduces the results of \citet{picogna2019dispersal}, where the \textsc{pluto} code was used.


\section{Diagnostics} \label{sec:diag}

In this section, we introduce quantities that we use for our diagnostics of the simulation output. Besides the X-ray luminosity, $\LX$, and as a standard measure for the relative magnitude of the magnetic effects, we use the ratio of thermal over magnetic pressure, that is, the dimensionless plasma $\beta$ parameter, $\betap \equiv 2\mu_0\,p/B^2$. We note that ---unless noted otherwise--- the quoted values characterises the midplane value of this quantity with respect to the initial disc model.


\subsection{Mass-loss and accretion rates}
\label{sec:massloss}

One of the most important and most commonly used characteristics for disc outflows is, of course, the wind mass loss rate:
\beq
  \left.\Mdotw \equiv 4\pi\int_{r_1}^{r_2}
  \rho v_{\theta} r \sin{\theta}\, {\rm d}r\,\right\vert_{\theta_a}\,,
  \label{eq:Mwind}
\eeq
where $r_1= 2\au, r_2= 22.5\au$ and $\theta_a$ the poloidal angle corresponding to the anchor point $z_a$ (an alternative to wind base, discussed in \Sec{he_cases}). Correspondingly, the mass accretion rate of the disc is:
\beq
  \Mdota = \left. 2\times \int_0^{\theta_d}
  2\pi\,\rho v_r r^2 \sin{\theta}\,{\rm d} \theta \, \right\vert_{r=r_i}\,,
  \label{eq:Macc}
\eeq
where $\theta_d$ marks the upper disc surface and $r_i\in[1.5, 2.5]\au$. The total mass accretion rate will then be computed from a time and radial average.

As, for instance, introduced in \citet{bai2013windI} or \citet{gressel2015global}, we also define the ``viscous'' accretion rate, $\dot{M}_\nu$, as the accretion rate that were to result if only turbulent transport of angular momentum was in effect:
\beq
  \left. \dot{M}_\nu = 2\Omega^{-1}\int_0^{\theta_d}
  3\pi \left( T^{\reyn}_{r\phi}
            + T^{\maxw}_{r\phi} \right)r\ {\rm d}\theta\,
  \right\vert_{r = r_i}\,,
  \label{eq:Mvisc}
\eeq
with $\;T^{\reyn}_{r\phi} \equiv \rho v_r(v_{\phi}-v_{K})\;$ and $\;T^{\maxw}_{r\phi} \equiv -B_r B_{\phi}/\mu_0\;$ the radial-azimuthal components of the Reynolds and Maxwell stress, respectively.

\subsection{Concepts related to magnetic disc winds}

Motivated by our desire to take a closer look on the fluid dynamics along a field line, we adopted some diagnostics previously employed by \citet{bai2017global,wang2019global,gressel2020global}, and study the forces acting on a fluid element along a magnetic poloidal field line:
\beq
  \frac{{\rm d}\vp}{{\rm d}t} = -\frac{1}{\rho}\frac{{\rm d}p}{{\rm d}s}
  + \left(\frac{v_\phi^2}{r}\frac{{\rm d}R}{{\rm d}s}
  - \frac{{\rm d}\Phi}{{\rm d}s}\right)
  - \frac{B_\phi}{\rho s\,\mu_0}\frac{\rm d}{{\rm d}s}\,\big( R\,B_\phi\big )\,,
  \label{eq:along_a_fl}
\eeq
where the first term in the \textit{rhs} represents the force due to gas pressure gradient, the middle two terms the effective centrifugal force, as the sum of the gravitational and the centrifugal force, and the forth term represents the force due to magnetic pressure gradient, and where $s$ is the coordinate along the field line.

We will also employ the wind invariants along a field line that are traditionally used to characterise a steady-state MHD wind \citep{blandford1982hydromagnetic}, that is for a vanishing net force across the field line. Note that these invariants are derived in the limit of ideal MHD, where the magnetic field lines are well coupled with the fluid. In this case, the invariants are:

The mass loading parameter, $\kappa$, which displays the amount of mass that is ejected by the wind
\beq
  \kappa \equiv \frac{\rho\,\vp}{\Bp}\,,
  \label{eq:kappa}
\eeq
the specific angular momentum, $\lambda$, which accounts for the angular momentum that is ejected by the wind
\beq
  \lambda \equiv R v_\phi - \frac{R B_\phi}{\mu_0 \kappa}\,,
  \label{eq:lambda}
\eeq
and the specific angular velocity, $\omega$,
\beq
  \omega \equiv \frac{v_\phi}{R} - \frac{\kappa B_\phi}{\rho R}\,,
  \label{eq:omega}
\eeq
where $\vp$ and $\Bp$ are the poloidal velocity and poloidal magnetic field strength, respectively, and $R$ is the cylindrical radius.

Related to the impact of non-ideal MHD on the appearance of disc winds, an important point of interest is to examine the relative strength and dominance of the three effects. To achieve this we use the dimensionless diffusivity ratios of each of the diffusive terms over the inductive term, as they are defined in \citet{wardle2012hall}, i.e.,
\beq
  \frac{O}{I} \equiv \frac{\etaO\Omega}{\vA^2}\,,\qquad
  \frac{A}{I} \equiv \frac{\Omega}{\gamma_i\rho_i}\,,\qquad
  \frac{H}{I} \equiv \frac{X}{2} = \frac{cB\Omega}{2\pi e n_e \vA^2}\,.
\eeq
Then, with the use of eqs.~(\ref{eq:diffusivities}), the ratios can be expressed in the general form $(\eta\Omega)/\vA^2$, that is, in terms of an (inverse) Elsasser number.


\begin{figure*}
 \begin{subfigure}{\linewidth}
      \includegraphics[width=0.95\linewidth]{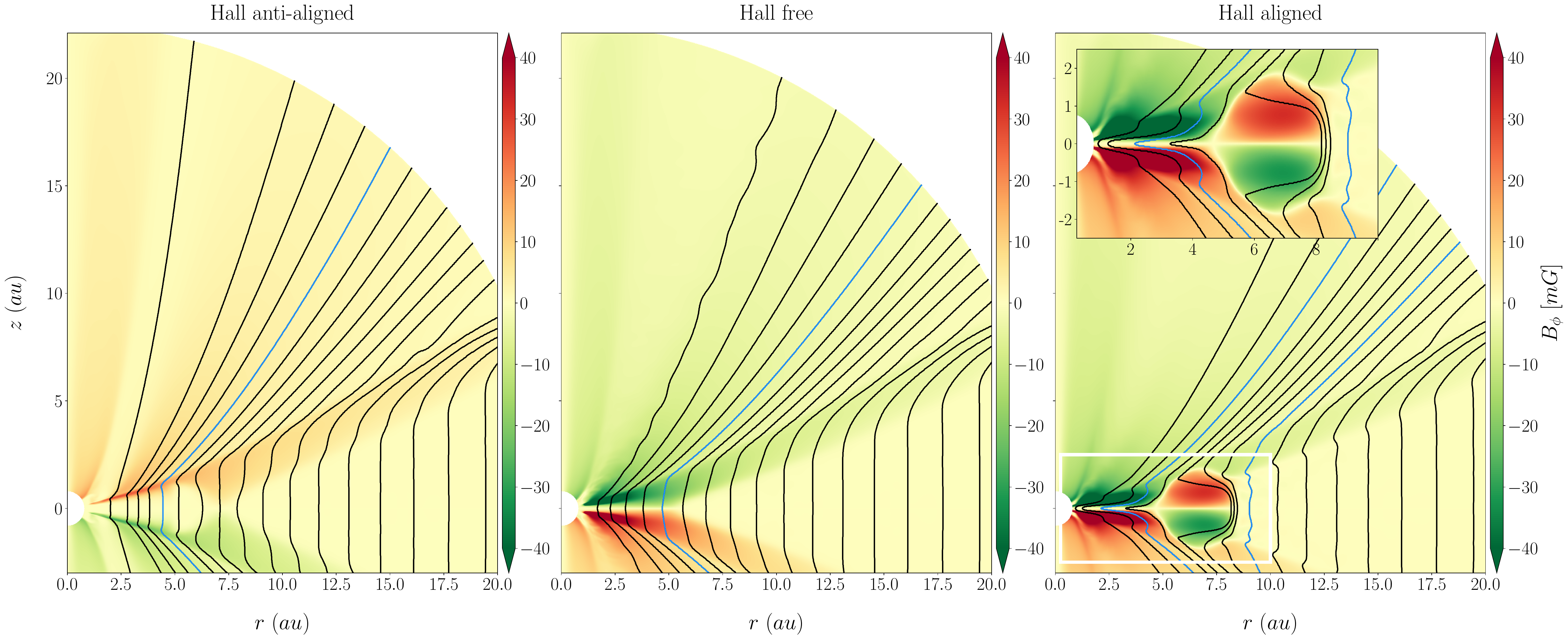}
      \caption{Toroidal magnetic field snapshots}
      \label{fig:bphi_mfl_all3}
    \hspace{1ex}
  \end{subfigure}
  \begin{subfigure}{\linewidth}
          \includegraphics[width=0.95\linewidth]{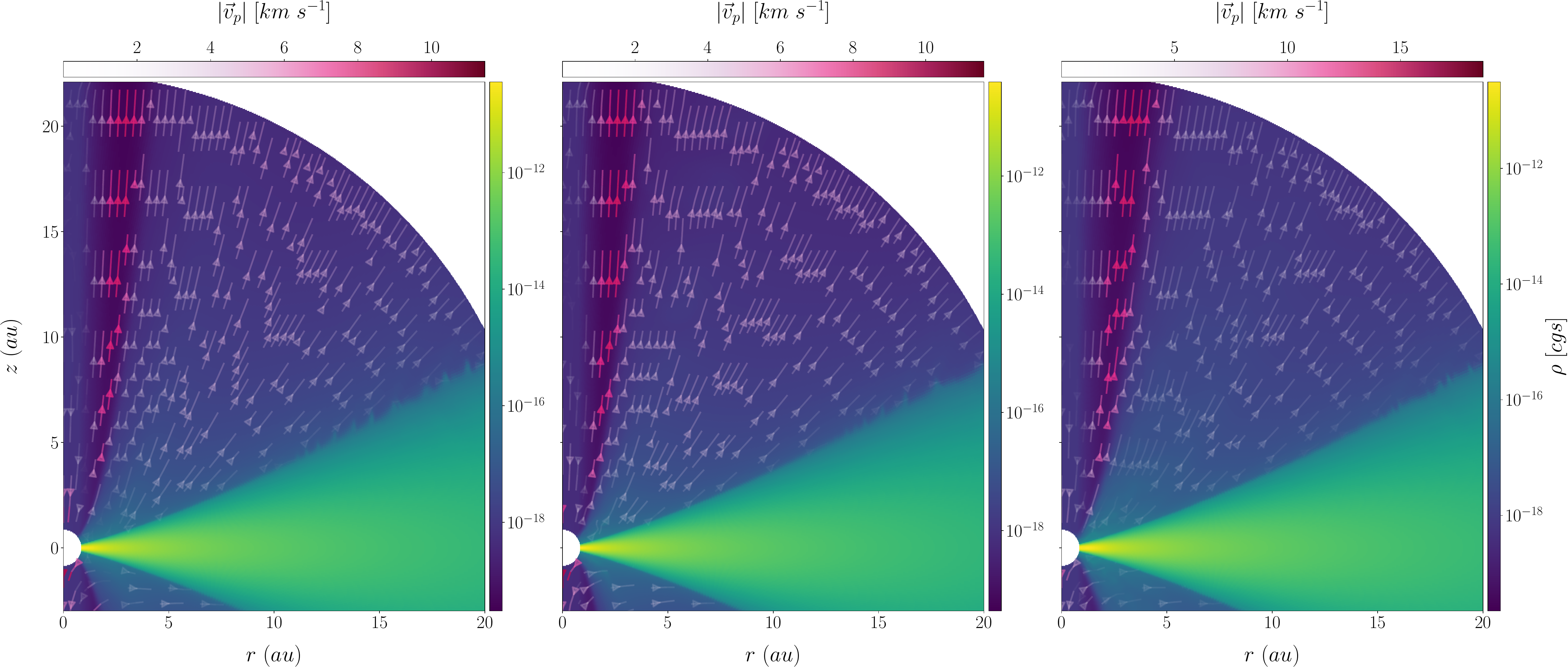}
          \caption{Density snapshots}
      \label{fig:rho_strm_all3}
    \hspace{1ex}
  \end{subfigure}
  \caption{Time averaged snapshots of the toroidal magnetic field with poloidal field lines (top row) and of the density with the poloidal velocity streamlines (bottom row) for the three different cases, the anti-aligned hall effect (\textit{HXRn5} - left), the hall free (\textit{AXRp5} - mid), and the aligned hall effect (\textit{HXRp5} - right).}
\end{figure*}


\section{Results} \label{sec:results}

We have performed a number of simulations that can roughly be grouped into the following categories: ~i) level of approximation with respect to micro-physics, ~ii) variation of the magnetic field strength (including the magnetic field orientation), ~iii) influence of the X-ray luminosity. A list of all simulations along with their labels and parameter values can be found in \Table{phys_dets}, where we also list other characteristic quantities that we have derived from the simulation results.


\subsection{The role of the Hall effect}\label{sec:he_cases}

\begin{figure*}
      \includegraphics[width=1\linewidth,]{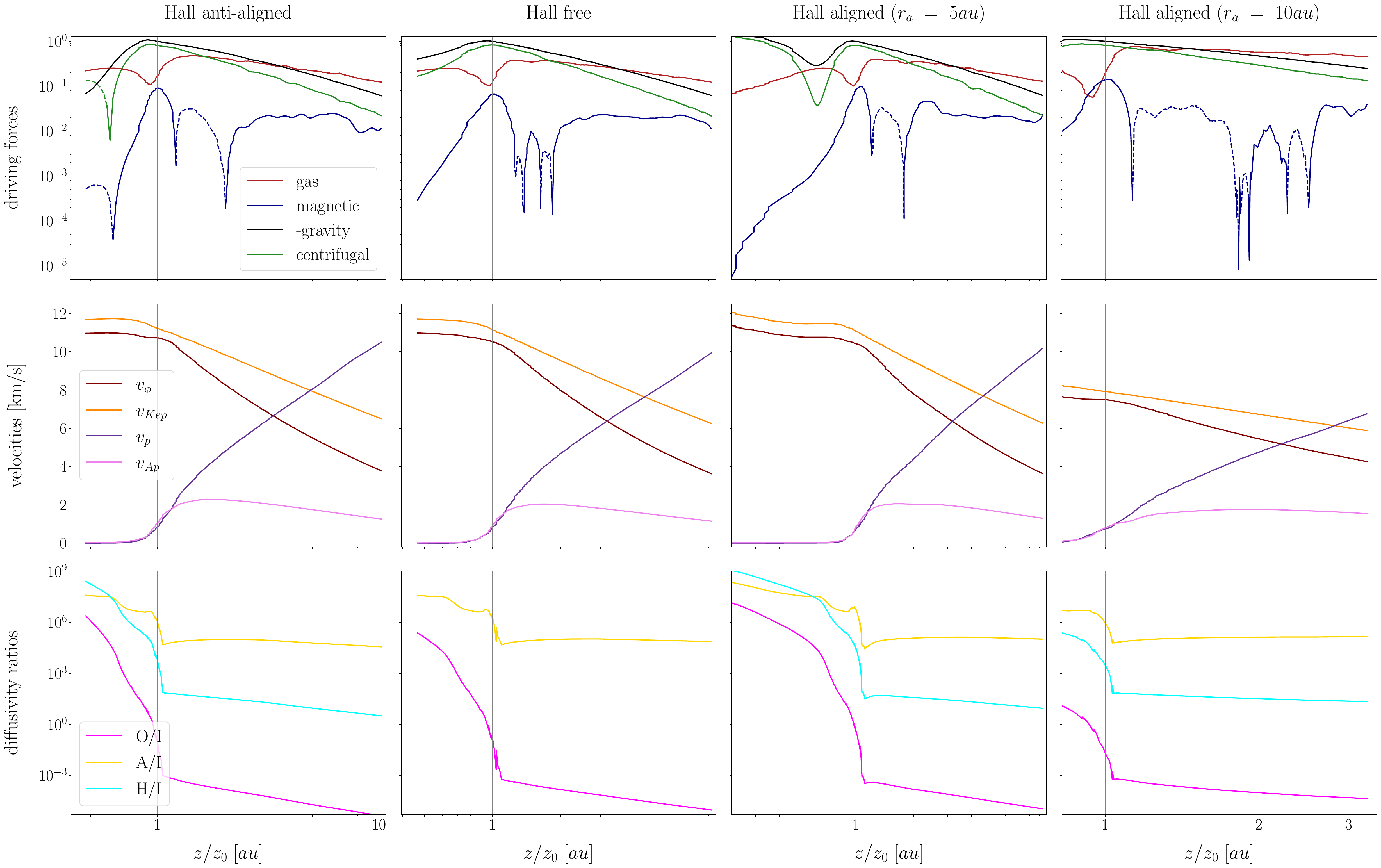}
      \captionsetup{width=\linewidth}
      \caption{Top row: various forces acted along a projected poloidal field line; middle row: several velocities along the field line; bottom row: the diffusivity ratios for the three different cases, Hall anti-aligned (left), Hall free (mid) and the aligned Hall effect (right). The dashed lines indicate negative values.}
      \label{fig:forces_along_fl_phys}
\end{figure*}

\begin{figure}
      \center\includegraphics[width=0.9\linewidth]{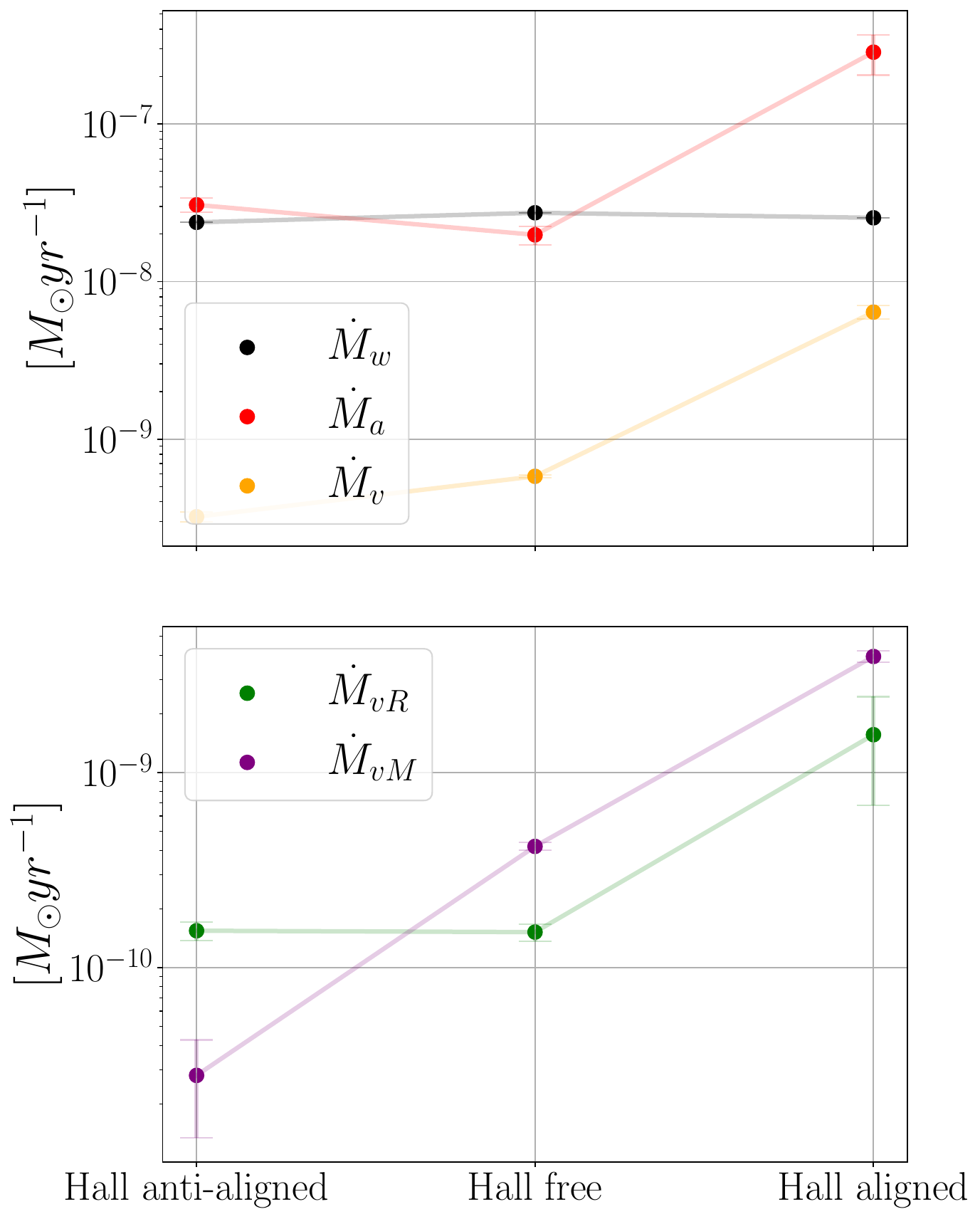}
      \captionsetup{width=0.6\linewidth}
      \caption{Total wind mass loss rate ($\Mdotw$), accretion accretion rate ($\Mdota$) and viscous accretion rates (i.e., $\Mdotv$, consisting of $\dot{M}_{vR}$ and $\dot{M}_{vM}$) of the three fiducial cases -- Hall free, anti-aligned and aligned HE. The error bars indicate the uncertainty in the time and volume averaging of the quantities.}
      \label{fig:M_scatter_phys}
\end{figure}

\begin{figure}
      \center\includegraphics[width=0.9\linewidth]{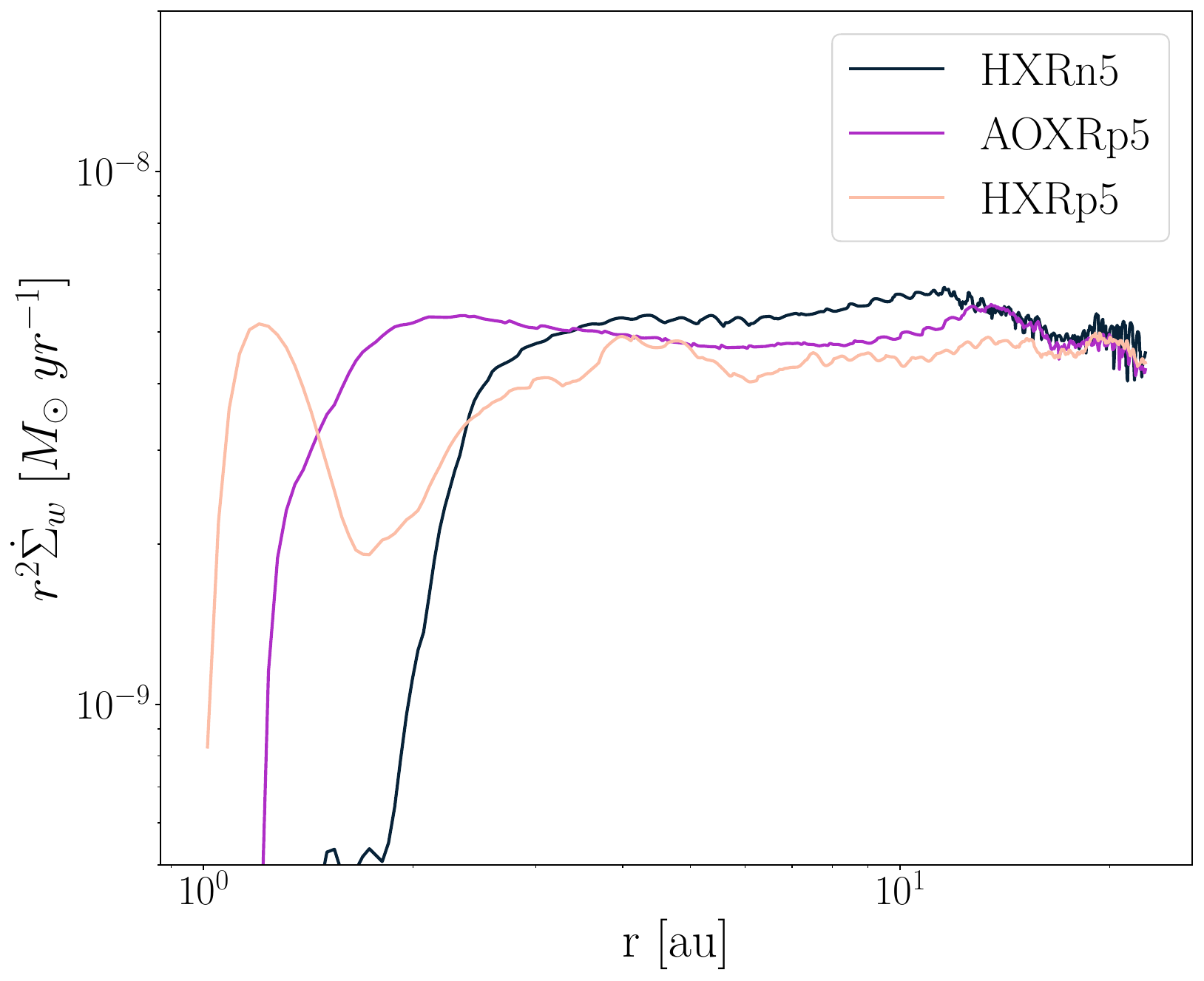}
      \captionsetup{width=0.6\linewidth}
      \caption{Surface mass-loss profile for the Hall anti-aligned \textit{(HXRn5)}, Hall free \textit{(AOXRp5)} and Hall aligned \textit{(HXRnp5)} cases.}
      \label{fig:sigma-dot-phys}
\end{figure}

\begin{figure*}
      \includegraphics[width=\linewidth]{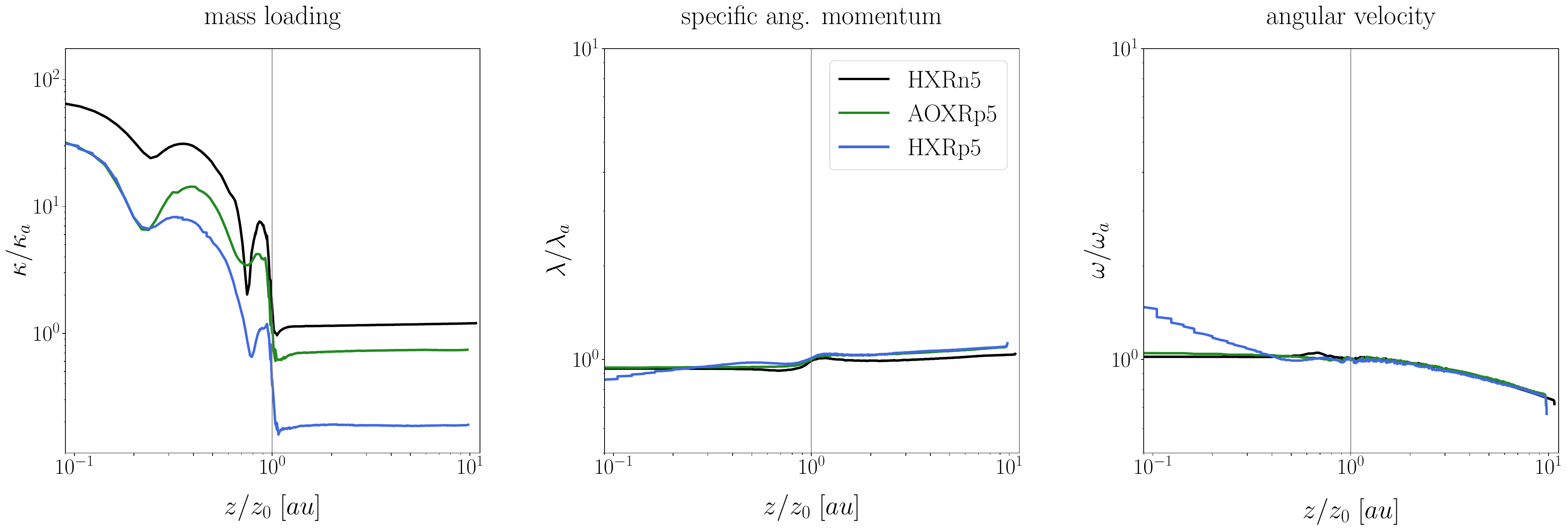}
      \captionsetup{width=0.6\linewidth}
      \caption{The MHD wind invariants of mass loading (left), specific angular momentum (mid) and specific angular velocity (right) for the anti- aligned, Hall free and aligned HE case - see \Tab{phys_dets} for the simulations' labels. The invariants have been normalised with their respective value at the anchor point.}
      \label{fig:invariants}
\end{figure*}


\begin{table}
\centering
\footnotesize
\begin{tabular}{lcccccccc}
\hline
\\[-6pt]
Simulation  & HE & $\betap$  & $\log{\LX}$ & $\Mdotw$    & $\Mdota$    \\[+3pt]
            &    &           & $\ergstb$   & $\Msunyrtb$ & $\Msunyrtb$ \\[+3pt]
\hline
\hline
HXRn5 & \chk & $\mns10^{5}$ & $30.3$ & $2.12\pm0.25$ & $3.07\pm0.32$ \\
AXRp5 & \nop & $10^{5}$     & $30.3$ & $2.48\pm0.27$ & $1.98\pm0.27$ \\
HXRp5 & \chk & $10^{5}$     & $30.3$ & $2.30\pm0.28$ & $28.6\pm8.10$ \\
\hline
HXRp3 & \chk & $10^{3}$     & $30.3$ & $6.89\pm0.79$ & $158.\pm28.0$ \\
HXRp4 & \chk & $10^{4}$     & $30.3$ & $3.73\pm0.38$ & $53.1\pm9.50$ \\
HXRp6 & \chk & $10^{6}$     & $30.3$ & $1.89\pm0.28$ & $22.5\pm6.50$ \\
HXRp7 & \chk & $10^{7}$     & $30.3$ & $1.63\pm0.30$ & $5.87\pm2.54$ \\
HXRp8 & \chk & $10^{8}$     & $30.3$ & $1.57\pm0.28$ & $2.27\pm0.83$ \\
HXRp9 & \chk & $10^{9}$     & $30.3$ & $1.57\pm0.28$ & $1.90\pm0.68$ \\
\hline
Lx293 & \chk & $10^{5}$     & $29.3$ & $1.09\pm0.03$ & $29.3\pm7.60$ \\
Lx298 & \chk & $10^{5}$     & $29.8$ & $1.55\pm0.04$ & $29.5\pm8.60$ \\
Lx308 & \chk & $10^{5}$     & $30.8$ & $6.64\pm0.89$ & $34.8\pm10.6$ \\
Lx313 & \chk & $10^{5}$     & $31.3$ & $19.1\pm2.90$ & $32.3\pm11.9$ \\
\hline
oHD\_Lx293  & --  & --      & $29.3$ & $0.12\pm0.03$ & $1.21\pm0.25$ \\
oHD\_Lx298  & --  & --      & $29.8$ & $0.47\pm0.08$ & $1.21\pm0.25$ \\
oHD\_Lx303  & --  & --      & $30.3$ & $1.50\pm0.29$ & $1.14\pm0.25$ \\
oHD\_Lx308  & --  & --      & $30.8$ & $5.07\pm0.96$ & $1.16\pm0.25$ \\
oHD\_Lx313  & --  & --      & $31.3$ & $16.5\pm3.10$ & $1.12\pm0.34$ \\
\hline
\end{tabular}
\caption{List of simulation runs, detailing whether the Hall Effect (HE) is included, the initial value of the plasma parameter $\betap$, the X-Ray luminosity, $\LX$, and the time averaged wind mass loss $(\Mdotw)$ and accretion rate $(\Mdota)$.}
\label{tab:phys_dets}
\end{table}


We have performed three fiducial runs, in order to directly compare how the inclusion of the HE influences the simulations. All three runs are taking into account Ohmic and Ambipolar diffusion, and they have an initial plasma parameter of $|\betap| = 10^5$ (in the disc mid-plane) and have X-ray luminosity of $\LX = 2\ee{30}\ergpersec$ (see \Table{phys_dets}, first four columns, for details). For two of these runs, we additionally include the HE, and focus on the overall orientation of the poloidal magnetic field, which becomes meaningful in this case.

The first of these simulations (termed `HXRn5`) has the initial magnetic field, $\B_0$, oriented anti-parallel to the disc's rotational axis (the ''anti-aligned'' case), while the second (termed `HXRp5`) has $\B_0$ parallel to $\hat{\mathbf{z}}$ (the ''aligned'' case). In the following, we compare and contrast the three cases according to key characteristics as well as the general morphology of the disc outflow.

The overall field topology of the three cases is shown in \Fig{bphi_mfl_all3}, where we present time averaged snapshots (from $t=75\yr$ to $90\yr$ with a $3\yr$ increment) of the toroidal magnetic field, and the projected poloidal field lines\footnote{We have exploited the enforced mirror symmetry at the mid-plane to expand the lower disc half, making the field polarity more directly visible.}.

We begin our discussion looking at the Hall-free case (\Fig{bphi_mfl_all3}, middle panel), where we observe ``field belts'' of positive (negative) $B_\phi$ at the innermost region of our disc. These belts are very similar to the ones found by \citet{gressel2020global} -- although they are somewhat smaller in our case, extending to $\sim 1\au$. We attribute this difference to some changes in the overall disc model (such as, the flaring index, or the radial surface density profile), that were implemented to more closely match the cases studied in \citet{picogna2019dispersal}. In the anti-aligned case (\Fig{bphi_mfl_all3}, left panel) ---despite the obvious reversal of the polarity (seen in the flipped sign of $B_\phi$)--- the overall topology of the field lines is quite similar compared with the Hall-free case. This especially holds true at large radii ($R>10\au$), where the field lines are threading the disc almost parallel to the rotational axis. At small radii ($R<4\au$), there is a marked difference, though. While the field lines are concave for the AD+OD model AXRp5 (that is, they emerge from the mid-plane with an outward pitch angle), the lines are essentially vertical for the anti-aligned model HXRn5. This can be explained by the radial component of the Hall velocity, $v_{{\rm H},R}\equiv\etaH/B\,\partial_z B_\phi$. For positive $\etaH$, this leads to an outward repulsion of magnetic flux in that case. This effect was first described in detail by \citet{bai2017hall}, see their Figure~1, right-hand panel. In the radial range $4\au < R < 8\au$, the field lines gradually take a convex appearance, that is, the repulsion due to the HE overcompensates the overall concave bending of the field lines. This in turn creates a slight back-and-forth wiggle in the field lines. As an overall consequence, the inclusion of the HE in the anti-aligned case, moreover leads to a general evacuation of $B_\phi$ from the midplane towards the wind base, creating pronounced current sheets for the region defined by $R<5\au$.

We now contrast the first two cases with that of the aligned Hall case shown in the right panel of \Fig{bphi_mfl_all3}. The most prominent feature of the magnetic topology is the sudden field reversal at about $R=5\au$, that we have magnified in the inset. Such a reversal was previously observed \citep[see fig.~5 in][]{gressel2015global}, and it may partly be a consequence of relic MRI ``channel modes'' with their characteristic localised radial field reversals \citep[e.g.][]{2010MNRAS.406..848L}. Such modes may appear at early times, before a strong $B_\phi$ emerges from the differential winding of the purely vertical initial field and subsequently quenches the growing MRI via AD \citep[see the discussion in ][]{gressel2015global}. Unlike in earlier Hall-free simulations, the inclusion of the aligned HE leads to a pronounced displacement of field lines. We note that a similar ``elbow'' shape has been observed by \cite{martel2022magnetised}. Although the authors considered the specific case of a jump in the surface density of the underlying disc (see Fig.~5 in their paper), the origin of this feature could well be related with ours.

The pronounced symmetry breaking observed in the inset of \Fig{bphi_mfl_all3}, highlights an important feature of Hall-MHD in the context of Keplerian rotation. As was already laid out by \citet{bai2017hall}, the aligned HE configuration tends to produce a pronounced inward ``pinch'' near the midplane, amplifying the overall concave wind topology. This is also clearly seen in our simulation -- but only for the region with $R<5\au$. So how can it be that we observe an outward ``repulsion'' of flux in the region $5\au < R < 8\au$, if such a behaviour is typically attributed to the anti-aligned case? The answer to this riddle lies in the sign of the azimuthal current, $J_\phi$. While \citet{bai2017hall} have generally assumed a concave bending of field lines for their argument, the polarity reversal also implies a \emph{convex} curvature in that region. As a consequence, this means that $J_\phi\,B_z$ picks up an additional minus sign compared to the now positive $B_z$ -- thus leading to the outward repulsion by the $\mathbf{J}\times\hat{\B}$ term in the induction equation. An important question for future investigations is whether the spontaneous appearance of convex curvature is likely to occur in real discs, or whether it is an artefact of the chosen initial configuration with vanishing azimuthal field.

In \Fig{rho_strm_all3}, we showcase snapshots of the averaged density, overlaid with poloidal velocity vectors. We note that in all three cases there is a stream with the highest velocity, corresponding to the least dense atmosphere region. There also exists some material falling towards the star near the axis. As discussed in \citet{gressel2020global}, this is likely a consequence of the excision of the inner sphere, so that there exist field lines in a cone around the axis that are not anchored in the disc. Luckily this seems to not influence the overall behaviour of our model, so at this point we chose to simply ignore it.

We now shift our focus to what is happening along a field line that is threading the wind (see the lines marked in pink colour in \Fig{bphi_mfl_all3}). Before we start, it has to be mentioned that defining a wind base proved quite tricky, since we had neither an FUV front \citep[as in][]{bai2017global,wang2019global} nor a clear transition from \textit{sub-} to \textit{super-} Keplerian rotation \citep{gressel2020global}. For this reason, we decided to treat the ``anchor point'' as the wind base, i.e., the point where the magnetic field line is emerging from the disc into the outflow.

In the upper row of \Fig{forces_along_fl_phys}, we demonstrate the forces acting along a field line, that is, the four terms described in \Eqn{along_a_fl}. Initially, we note that the effective centrifugal force (i.e., the sum of the gravity and centrifugal terms, $F_{\rm eff}$) is always negative; \citep[as observed in][]{bai2017global,wang2019global}, but in contrast to \citep{gressel2020global}. This implies that co-rotation is never enforced and thus cannot drive the centrifugal launching of the outflow, necessary for the production of a magneto-centrifugal wind \citep{blandford1982hydromagnetic}. This is also confirmed by looking at the poloidal Alfv{\'e}n velocity $v_{\rm Ap}$ in \Fig{forces_along_fl_phys}; as pointed out in \citet{Bai2016Thermal}, when $v_{\rm Ap}<v_{K}$ then $\Feff<0$. At the same time, we notice a simultaneous dip on the force due to gas pressure ($F_{\rm gas}$) and a peak on the force due to magnetic pressure ($F_{\rm mag}$). In addition, the dip and the peak that is seen in these two quantities, respectively, appear to have similar (normalised) values. This would imply that at the anchor point, the contribution for the wind launching comes in approximately equal parts from $F_{\rm gas}$ and $F_{\rm mag}$. However, after the anchor point $(z/z_0 \gtrsim 1.1\au)$, $F_{\rm gas}\geqslant F_{\rm mag}$, a clear indication that the wind is actually thermally sustained.

In terms of the prevalence of non-ideal effects (bottom row of \Fig{forces_along_fl_phys}), everything aligns with our anticipated outcomes. Ambipolar Diffusion takes the lead in the majority of regions, with the exception being smaller radii, where the Hall Effect is the dominating instead. The reason the Ohmic Resistivity isn't the dominant factor lies in the fact that our simulation domain starts at $0.75\au$, which is not as close to the star for OR to play a significant role.

We also have to mention that for the aligned case we followed two poloidal field lines, that were located to the left ($r = 5\au$) and right ($r = 10\au$) from the magnetically reversed region. No meaningful differences were found between the two cases, indicating that this field reversal affects only the behaviour of the field lines inside the disc rather than the outflow itself.

With \Figure{M_scatter_phys}, we now turn our attention to one of the central outcomes of our simulations, that is, we take a look at the various accretion- and mass-loss rates (see \Sec{massloss} for the definition of the various expressions). Besides the disc accretion rate $\Mdota$ and the wind mass--loss rate $\Mdotw$ ---see Eqns. (\ref{eq:Macc}) and (\ref{eq:Mwind}), respectively--- , we have also calculated an approximation for the total ``viscous'' accretion rate $\Mdotv$ ---see \Eqn{Mvisc}--- which we split into kinematic and magnetic contributions $(\dot{M}_{\rm vR}$ and $\dot{M}_{\rm vM})$, arising from the Reynolds and Maxwell stresses, respectively.

One can immediately notice that the wind mass loss rate $\Mdotw$ is nearly invariant for the three cases considered. This is not surprising, since the nature of the outflow is thermal in all three cases (see \Fig{forces_along_fl_phys}, above). Additionally we see that the mass accretion rate is approximately 2 orders of magnitude larger than the viscous accretion rate. From this we can infer that the main source of mass accretion is the MHD wind - as pointed out in \citet{bai2017global} we can have angular momentum removal from an MHD wind that is not necessarily magnetocentrifugal.

Furthermore, we can very nicely observe the duality of the HE in the laminar accretion due to Maxwell's stress, $\dot{M}_{vM}$. As mentioned before, when $\hat{z}\vec{\Omega}>0$ then the HE is enhancing any buckling of the magnetic field lines in the midplane and thus enhances the horizontal magnetic field strength responsible for the $T^{\maxw}_{R\phi}$ contribution. In the opposite case, the radial component of the field line gets reduced by the outward push, conversely resulting in a reduced magnetic contribution. This translates directly to what we observed in the bottom panel of \Fig{M_scatter_phys}, that is, $\dot{M}_{vM}$ has the minimum value in the anti-aligned case $(\sim 2\ee{-11}\Msunyr)$ and the maximum for the aligned HE case ($\sim 4\ee{-9}\Msunyr$). At the same time we see an increase in the Reynolds stress for the Hall aligned case. This could potentially be a result of the enhanced magnetic field topology from the HE - although the exact details are currently unclear.

Analyzing the surface mass-loss profiles in \Figure{sigma-dot-phys} reveals an interesting point regarding the HE. In the anti-aligned case, the wind first originates at $\simeq 2.5\au$, while in the Hall-free and Hall-aligned cases, it already begins at $\simeq 1.5\au$ and $\simeq 1\au$, respectively. This difference likely stems from variations in magnetic field distribution in the inner region. As we have pointed out, the Hall anti-aligned exhibits a lack of $B_{\phi}$ near the midplane, leading to wind launching only becoming efficient at larger radii. Conversely, the Hall aligned displays a concentration of strong $B_{\phi}$ within $\sim 5au$ of the midplane, resulting in wind launch from smaller radii -- albeit with a pronounced ``trough'' at $\simeq 1.5\au$, that will require some more detailed examination.

Finally, we turn our attention to the wind invariants, as depicted in \Fig{invariants} -- see Eqns. (\ref{eq:kappa})--(\ref{eq:omega}) for their respective definitions. As expected the quantities for $z/z_0>1$ are fairly constant along the field line, re-affirming that in the X-ray heated disc atmosphere we are fairly close to the ideal MHD limit, despite AD dominating below the anchor point.


\subsection{Sensitivity and dependence on key parameters}\label{sec:param}

Starting from the fiducial scenario discussed above, we now venture into understanding how the key outcomes of our models depend on the various underlying properties and assumptions. To this end, we have done an extensive parameter study, where we have investigated the dependency of the wind mass loss rate with the initial plasma parameter, $\betap$, and with the star's X-ray luminosity $\LX$. Regarding the last point, in  \citet{ercolano2021dispersal}, an extended temperature parametrisation was introduced, which prescribed different temperature parameters depending on $\LX$ (see \Sec{xray_heating}). In this part of our work, this is ignored, since an additional goal of this section is to examine how our MHD models compare with \citet{picogna2019dispersal}. Hence the temperature parameters have remained the same as for $\LX = 2\times10^{30}~\ergpersec$, and only the value of $\LX$ is varied.

\begin{figure}
      \center\includegraphics[width=0.9\linewidth]{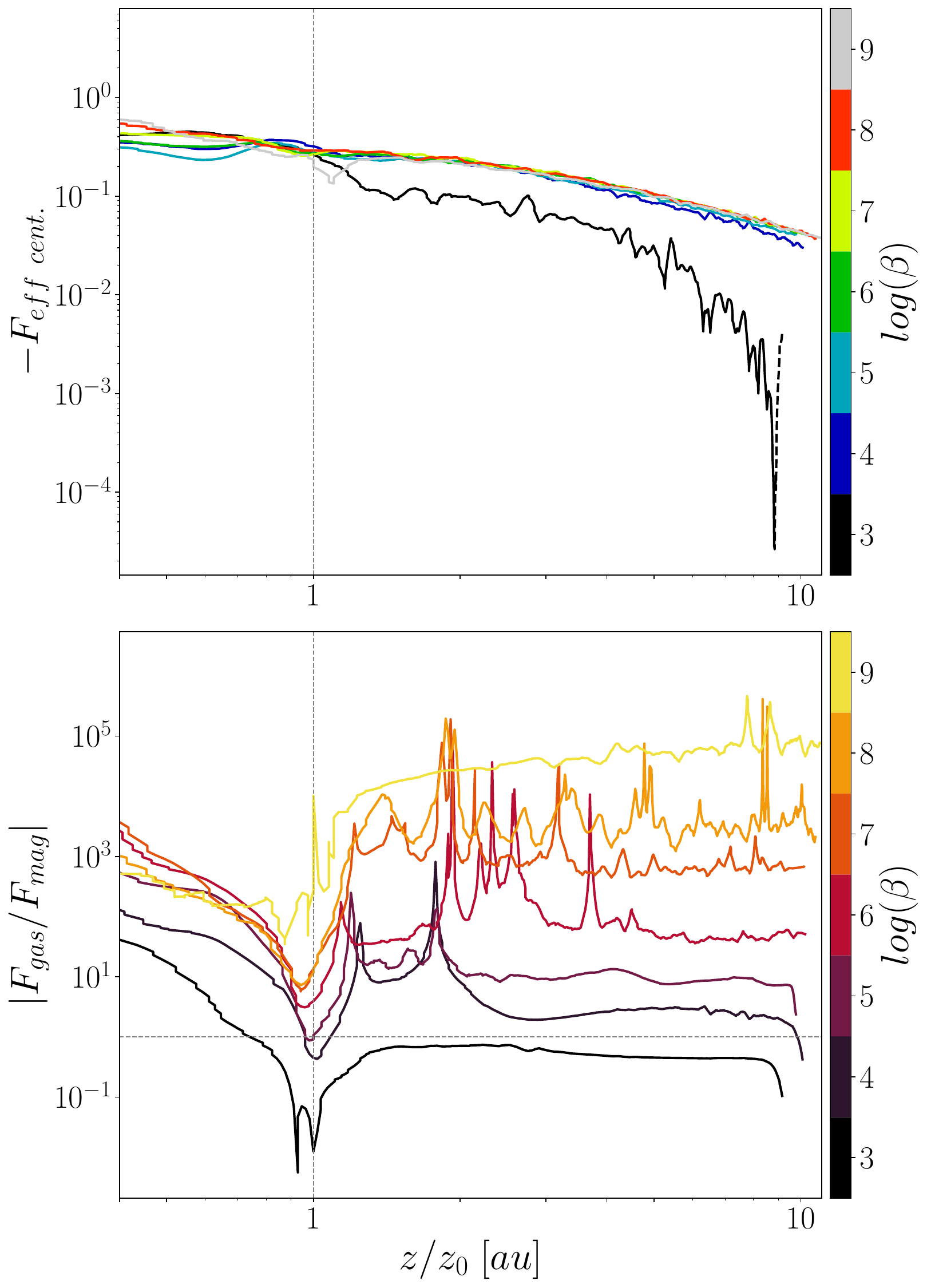}
      \captionsetup{width=0.6\linewidth}
      \caption{Top panel: The effective centrifugal force along a poloidal field line; bottom panel: the absolute ratio of the gas pressure gradient and the magnetic pressure gradient along the same field line, with $\LX = 2\times10^{30}\ergpersec$ and several values of the plasma parameter $\betap$. A moving average has been applied for the easier separation of the lines; the dashed parts in the line indicate negative values.}
      \label{fig:ratios_beta_par5}
\end{figure}


\subsubsection{Dependence on the plasma parameter}\label{sec:plasma_parameter}

The goal of this section is to investigate the influence of the overall magnetic field strength onto the nature of the wind -- and additionally onto the mass loss and mass accretion rates. For this, we choose the Hall aligned case, as the one showcasing the most ``extreme'' characteristics of the HE. For the set of simulations presented here, we kept the disc model and X-ray luminosity the same as our fiducial run, while varying the plasma parameter $\betap$ as $\logb = 3, 4, 5, \dots, 9$.

\begin{figure}
      \includegraphics[width=0.9\linewidth]{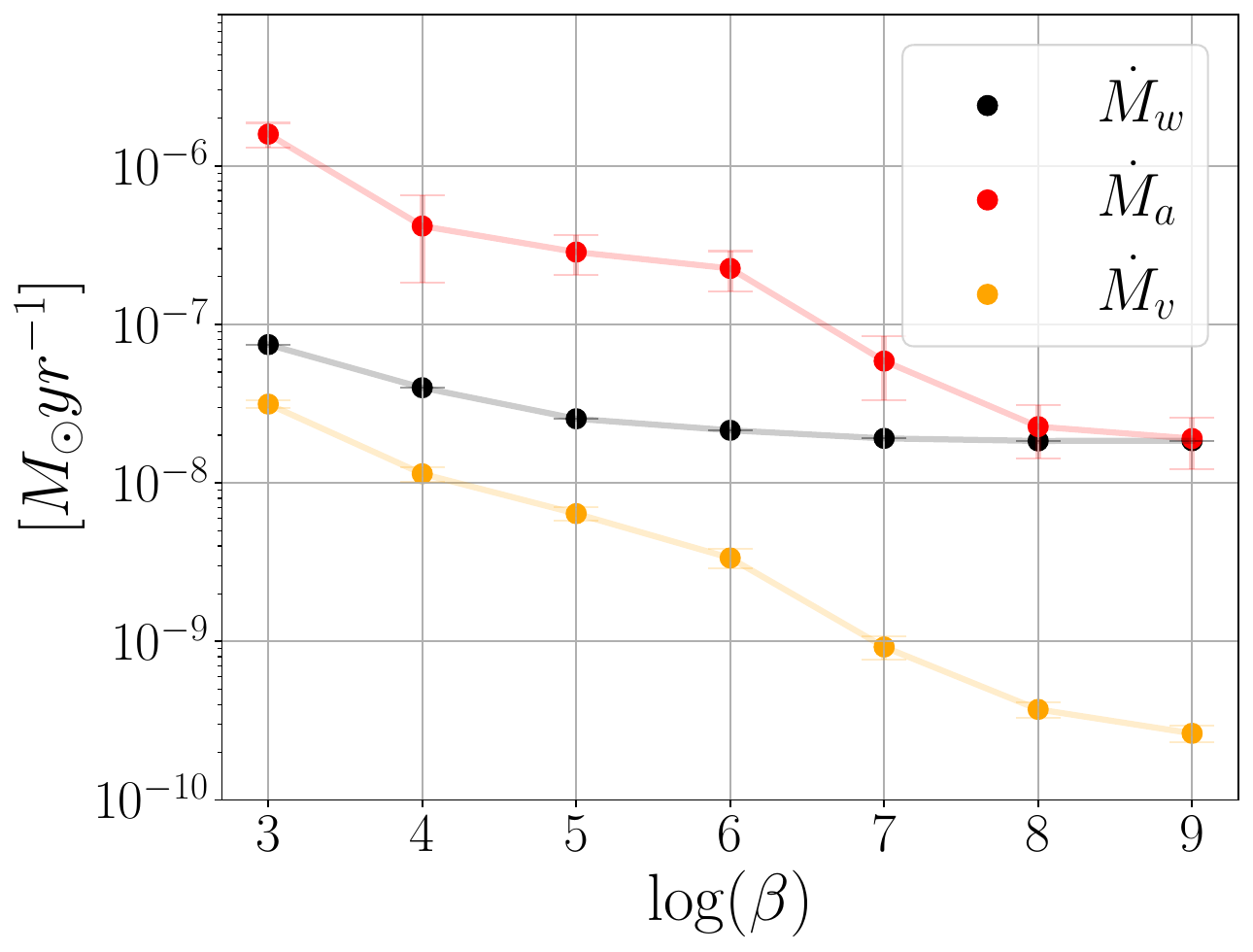}
      \captionsetup{width=0.6\linewidth}
      \caption{Trends of the total wind mass loss rate ($\Mdotw$), mass accretion rate ($\Mdota$) and viscous accretion rate ($\Mdotv$). The error bars indicate the uncertainty of the measurements.}
      \label{fig:M_scatterM_vs_beta}
\end{figure}

\begin{figure}
      \includegraphics[width=0.9\linewidth]{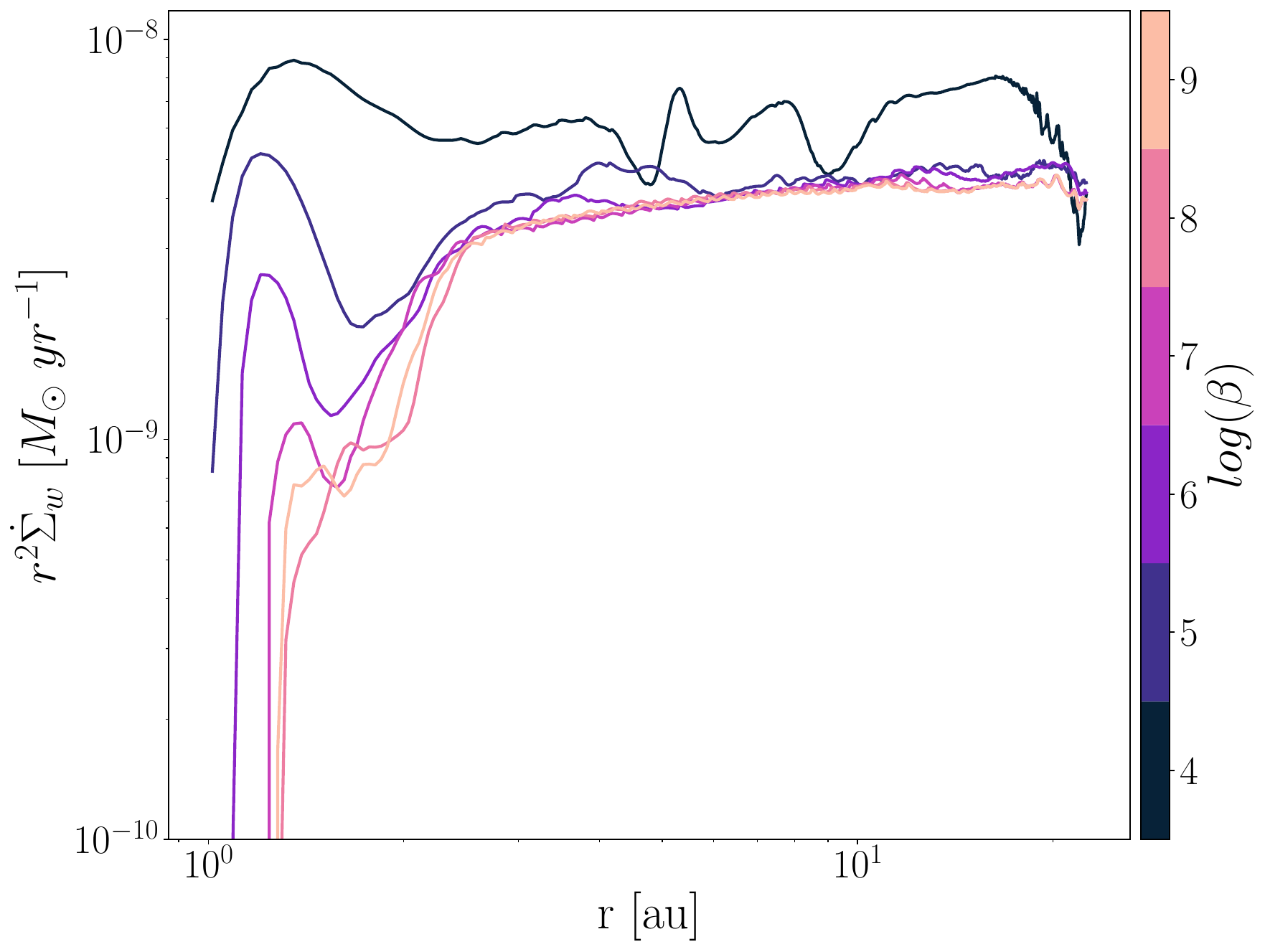}
      \captionsetup{width=0.6\linewidth}
      \caption{Surface mass loss profiles for runs with $\loglx = 30.3$ and $\logb = 4, 5, 6,..., 9$}
      \label{fig:sigma-dot-beta}
\end{figure}

In \Fig{ratios_beta_par5}, we demonstrate the effective centrifugal force, $\Feff$ ---i.e., the sum of the second and third \textit{rhs} term in eq. \Eqn{along_a_fl}--- as well as the absolute ratio of the force due to gas pressure gradient over the one due to magnetic pressure gradient, $\fgfm$. It seems that while $\log{\betap}>3$, $\Feff$ remains negative and essentially unaffected by the change of the magnetic field strength; the wind is not magneto-centrifugal for any of the studied values of $\betap$. For $\logb = 3$ at $z\sim 8z_0$, $\Feff$ actually reaches positive values. Since this is not happening anywhere near the wind launching region it is quite unlikely that this could affect the nature of the wind.

Additionally, we observe that there are some clear trends in the force ratio. Predictably, as the magnetic field strength weakens (i.e., for increasing $\betap$), the ratio increases as well since $F_{\rm gas}$ is expected to remain more or less the same. What is interesting is that there is a notable dip around the anchor point for all cases. In a more detailed look, we note that for $\logb = 6, 7, 8, 9$, we find $\fgfm>1$ along the entirety of the field line. On the other hand, for $\logb = 4, 5$, the dip around the anchor point reaches values of $\fgfm<1$, while for $\log{\betap} = 3$, the ratio is below unity for $z/z_0>0.7\au$. We infer that for $\log{\betap}\gtrsim 6$, the magnetic field is quite weak without being able to drive any outflow -- the wind is essentially photoevaporative in nature. For $\log{\betap}\lesssim 5$, the wind is probably launched due to the magnetic pressure gradient, but unless the magnetic field strength is rather unrealistically high (i.e., $\log{\betap} \simeq 3$) , it is sustained mainly by thermal pressure.

Compiling these findings in \Fig{M_scatterM_vs_beta}, we note that for the cases where the wind was dominantly magnetically driven, we get a slightly higher wind mass loss rate than the ones dominated from photoevaporation (see \Table{phys_dets} for the detailed values). Furthermore, for the predominantly thermal outflows (i.e., $\logb = 6, 7, 8, 9$), the wind mass loss rate converges to typical value of $\sim 10^{-8}\Msunyr$, that has been established in works focused on purely hydrodynamical models of discs with X-ray heating \citep{2008ApJ...688..398E,2009ApJ...699.1639E,owen2010radiation}.

This can additionally be inferred by looking at \Figure{sigma-dot-beta}, focussing on the region $r>2.5\au$. However, another point emerges when looking at smaller radii; first, in all the lines we see a characteristic \textit{``bump and dip''} profile that seems to be characteristic for the aligned HE case. Increasing the magnetic field (decreasing $\beta$) causes the wind to be launched from progressively smaller radii and also enhances the overall mass loss--rate.

Furthermore, as previously noted, the observed accretion rates appear to be primarily driven by the MHD wind but also, to a lesser extent, influenced by laminar stress


\subsubsection{Dependency on the X-ray luminosity}\label{sec:xraylum}

We now explore how the wind is affected by changing the X-ray luminosity while keeping the magnetic field strength constant (at $\betap = 10^5$). As before, we focus on the Hall aligned case.

\begin{figure}
      \includegraphics[width=0.9\linewidth]{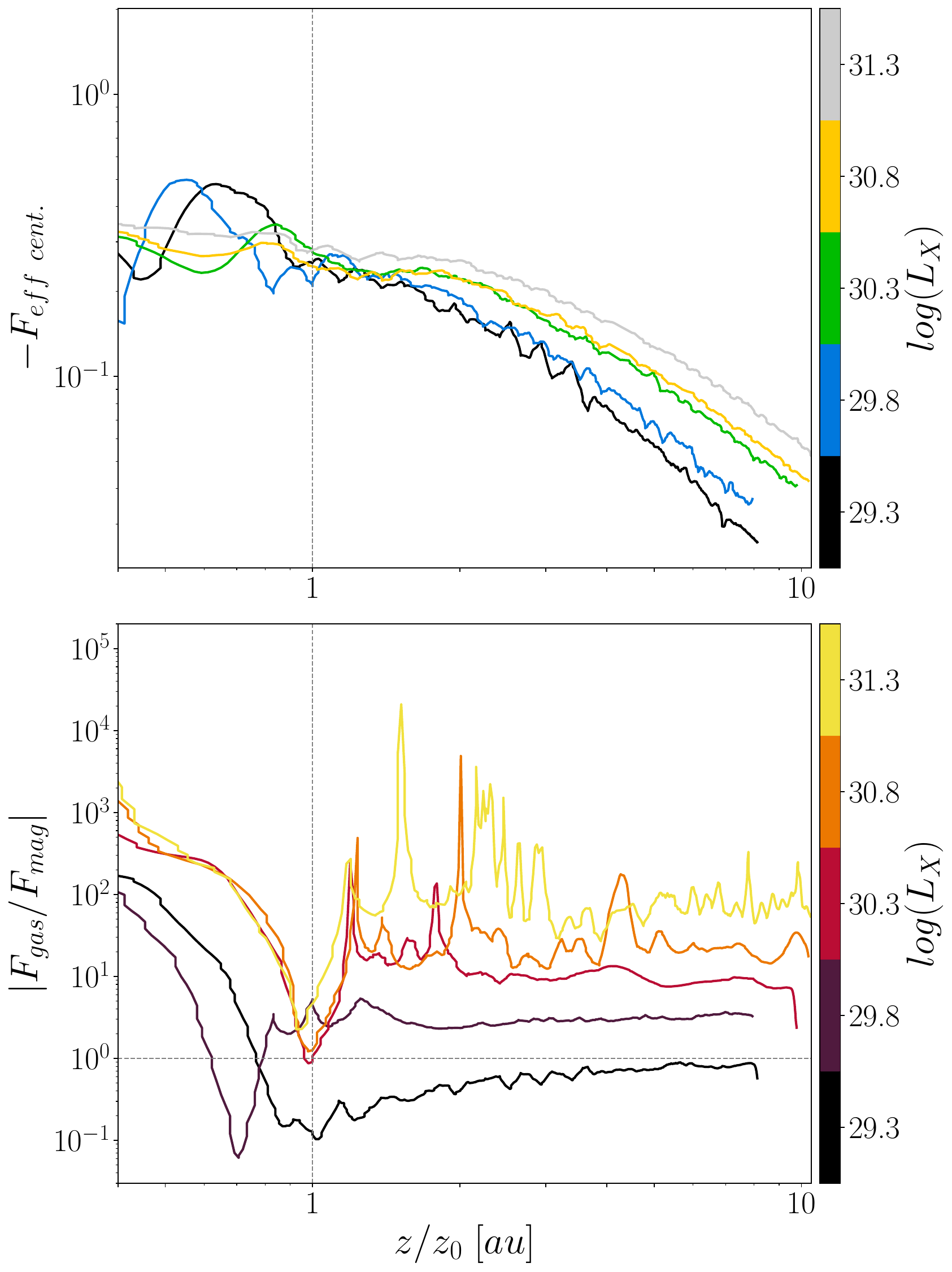}
      \captionsetup{width=0.5\linewidth}
      \caption{Top panel: The effective centrifugal force along a projected poloidal field line; bottom row: the absolute ratio of the gas pressure gradient and the magnetic pressure gradient along the same field line, with $\LX = 2\times10^{30}\ergpersec$ and for several values of the plasma parameter $\betap$.}
      \label{fig:ratios_Lx_par5}
\end{figure}

\begin{figure}
      \includegraphics[width=0.9\linewidth]{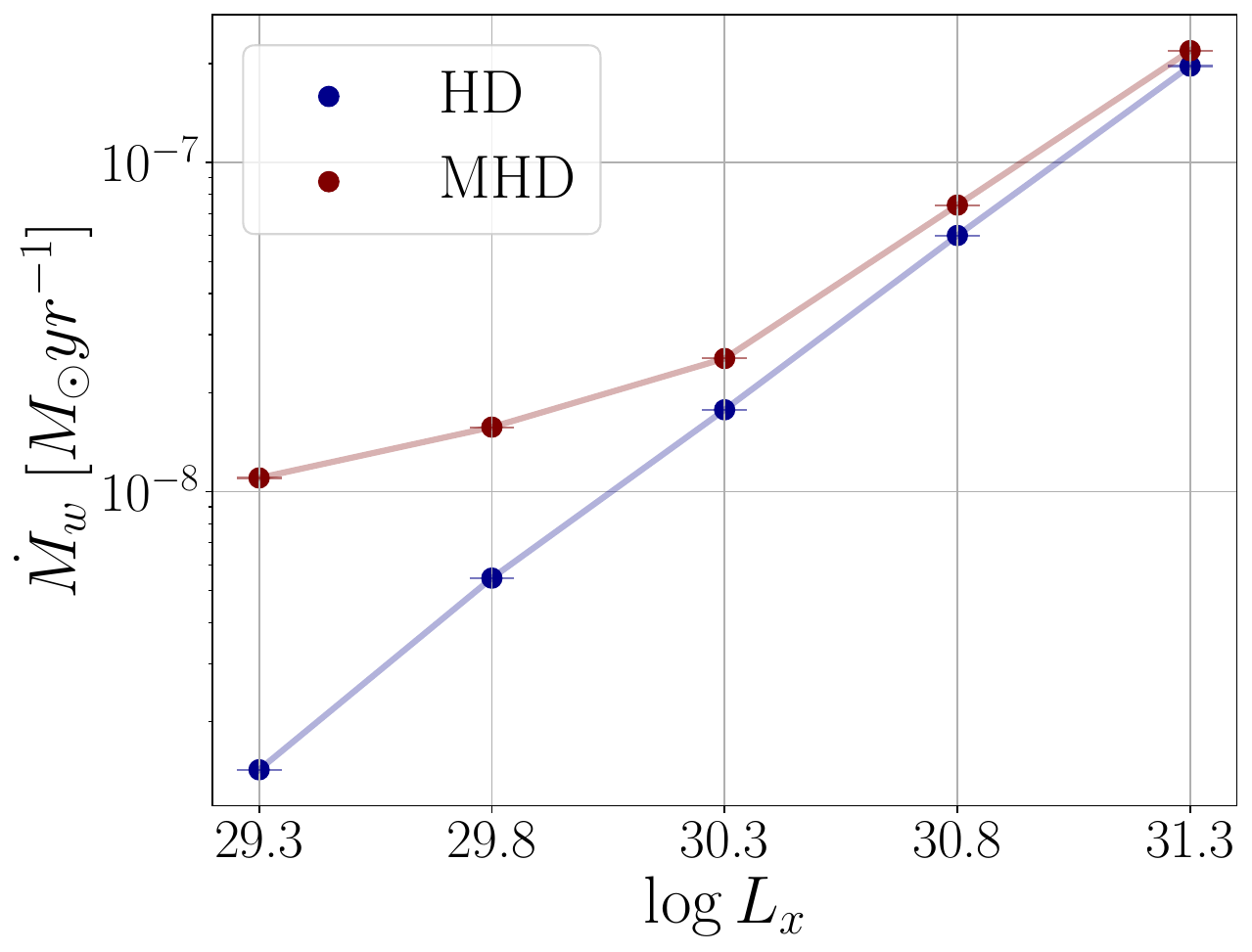}
      \captionsetup{width=0.5\linewidth}
      \caption{Dependence of the wind mass loss rates ($\Mdotw$) for runs with several $L_X$ values for a set of runs with MHD and one with only hydro.}
      \label{fig:Mw_MHD_vs_HD}
\end{figure}

Similar to the case of the $\beta$ dependence (see \Fig{ratios_beta_par5} above), in \Fig{ratios_Lx_par5} on the top panel we now plot the effective centrifugal force as a function of X-ray luminosity. $\Feff$ is negative for all cases, along the entire poloidal field line; we can thus safely assume that we do not observe a true MCW -- no matter how low the value of $\LX$. This is hardly surprising since the magnetic field strength is unaltered. In the bottom panel of \Fig{ratios_beta_par5}, as before we plot the absolute ratio of the forces due to the gas and magnetic pressure gradient. The picture we see is quite similar with the one in \Sec{plasma_parameter}. We again observe the sudden dip of the ratio in the area around the anchor point. We also get a clear trend: as $\LX$ is decreased, the ratio is also decreasing. This is expected since with lower $\LX$ the thermal pressure developed in the disc's surface is weaker and cannot contribute as much to the wind's acceleration.

Turning our attention to the mass loss and accretion rates, we introduce a new set of simulations. Given that for $\logb = 5$ we find a mainly photoevaporative wind, we decided to compare the MHD simulations with hydrodynamic ones. Removing the magnetic field from our initial model, we conducted a total of six runs with $\LX$ values identical to those in the corresponding MHD runs.

\begin{figure*}
      \includegraphics[width=0.9\linewidth]{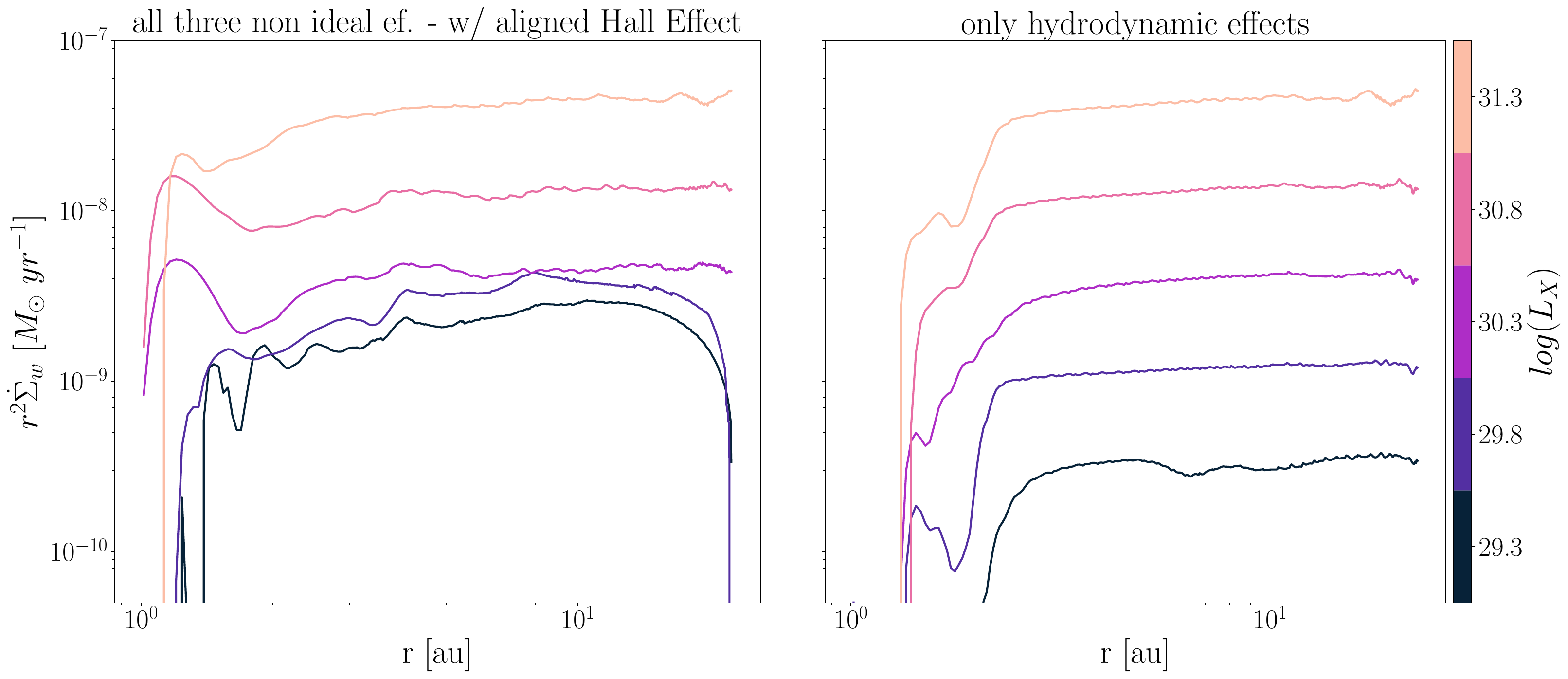}
      \captionsetup{width=0.5\linewidth}
      \caption{Surface mass-loss profiles for the MHD (left) and the HD (right) runs for different $L_X$ values.}
      \label{fig:sigma-dot-Lx}
\end{figure*}

In Fig. \ref{fig:Mw_MHD_vs_HD} we plot the wind mass loss rate for the two cases (MHD vs HD). In both we see a clear increasing trend with increasing $\LX$. However it is interesting to point out that the difference of $\Mdotw$ between the MHD and HD cases is decreasing with increasing$\LX$. This implies that the magnetic field strength does indeed contribute to $\Mdotw$, albeit only discernibly so for low enough X-ray luminosity values ($\loglx \lesssim 30.8$). For values higher than that, the measured wind mass loss rates are predominantly due to the photo-evaporative mechanism.

Additionally, as before, we also compare the surface mass-loss profiles for the MHD and HD runs in \Figure{sigma-dot-Lx}. Looking again at the region with $r<2.5\au$, we can clearly see the progressively increasing magnetic contribution to the outflow, implying that discs with significant magnetic flux are capable of launching winds from smaller radii than their purely hydrodynamical counterparts.

\section{Conclusions} \label{sec:concl}

We have successfully performed simulations with all three non-ideal MHD effects while simultaneously employing the X-ray temperature parametrisation that was introduced in \cite{picogna2019dispersal}. The central goal of our study was to examine if and how a photoevaporative wind is altered when taking into account non-ideal MHD effects.

Initially, we have focused on studying specifically the consequences of including the Hall Effect. To this end, we took into account three cases, the anti-aligned HE (when $\hat{z}\cdot\vec{B}<0$), Hall free (including only Ohmic Resistivity and Ambipolar Diffusion) and aligned HE ($\hat{z}\cdot\vec{B}>0$). Starting with the poloidal field lines, we note that in the Hall anti-aligned case they are vertically threading the disc. An opposite picture presents itself in the Hall aligned case. There exists a sudden field reversal, at $r\in(5,8)\au$ and for $r<5\au$ we have prominent inward pinching of the field lines, while for $r>5\au$ they are bend outwards. This is directly connected to the nature of the aligned Hall effect to amplify and pronounce any existing curvature of the field lines. A similar behaviour was previously observed by \citet{bai2017global}. Whether the field reversal in our simulations is a relic of the initial configuration (or not) is something to still be figured out. In all the above cases, though, the wind is magnetically assisted near the launching region, but thermally sustained beyond that. Overall, the wind mass loss rate remains roughly constant at around $\Mdotw = (2.3\pm 0.2)\ee{-8}$, suggesting that these differences in microphysics do not play a major role in the wind's launching process.

Next, we have focused on how the character of the wind changes while altering $\betap$ and $\LX$, respectively. Unsurprisingly, the higher the magnetic field strength, the more magnetically assisted the wind becomes. The limiting value for obtaining a magnetically dominated wind seems to be $\logb = 3$ while $\logb = 6$ is the threshold for a predominantly thermally driven wind. The equivalent trend is observed in the X-ray luminosity, the stronger it becomes, the more thermal the wind launching. The critical values, in this case, appear to be $\loglx = 29.3$ and $\loglx = 30.8$ for the magnetically and thermally dominated limits, respectively.

All the models are able to produce winds with mass loss rates in the expected range of $10^{-8}$ to $10^{-7}\Msunyr$. A notable observation is that when we have a strong enough magnetic field (i.e., $\logb\leq4$), we measure mass loss rates a factor of about ten larger than the purely photoevaporative ones. Additionally, if $\LX$ is strong enough (i.e., $\loglx \geq 30.3$), then the mass loss rates predicted from purely hydrodynamic models begin to match the ones predicted from MHD models. We can assume that there is a combination of magnetic field strength and X-ray luminosity that would produce a mass loss rate that could be approximated by an ``hydro-only'' model.

According to the findings presented in \citet{rodenkirch2020global}, where only OR (but not AD or the HE) was included, the authors proposed that for $\betap\geq 10^7$, the wind mass loss rates converge towards the rates of pure photoevaporation. However, based on \Fig{M_scatterM_vs_beta}, we find that in our simulations this happens even sooner, for $\betap\geq 10^5$. Their model only included Ohmic Resistivity, hence this could suggest that by including all three non-ideal MHD effects, photoevaporation already dominates the wind launching for even stronger magnetic fields than previously assumed. At the same time, the mass accretion rate clearly increases with the magnetic field strength, with a convergence to a stable rate appearing for $\betap \geq 10^8$.

As far as $\Mdota$ is concerned, we find that most of it is due to the MHD wind, which is driven by a magnetic pressure gradient, in our case. Additionally, a small contribution is required, stemming from the combined stress tensor, that is, effected from kinetic (via $T_{R\phi}^{\reyn})$ and magnetic (via $T_{R\phi}^{\maxw}$) contributions. While we can link the origin of an enhanced ``laminar'' magnetic stress to the aligned HE, the presence of a higher Reynolds stress in this case is still unaccounted for.

Last but not least, we have managed to unambiguously identify the predominant wind launching areas for the magnetic and the photoevaporative mechanism, respectively. In agreement with theoretical predictions by \citet{ercolano2017dispersal}, we find that the magnetic winds are typically launched from the very inner radii of the disc (e.g., $r\leq 1.5\au$ in the aligned HE case), while the photoevaporative outflow is only possible beyond the gravitational radius, which is clearly further away from the central star in our models.

\section*{Acknowledgements}

We thank C.~Dullemond, and P.~Rodenkirch for useful discussions. This work used the \nirv MHD code version 3.8, developed by Udo Ziegler at the Leibniz-Institut f{\"u}r Astrophysik Potsdam (AIP). Computations were performed on the \textsc{taurus} node of the local AIP cluster. We acknowledge support of the DFG (German Research Foundation), Research Unit `Transition discs`, funding ID 325594231. This work was co-funded\,\footnote{Views and opinions expressed are however those of the author(s) only and do not necessarily reflect those of the European Union or the European Research Council. Neither the European Union nor the granting authority can be held responsible for them.} by the European Union (ERC-CoG, \textsc{Epoch-of-Taurus}, No. 101043302).


\section*{Data Availability}

The simulation data produced in our runs as well as details concerning the numerical setup will be shared on reasonable request to the corresponding author.

\bibliographystyle{mnras}
\bibliography{ref}

\begin{thebibliography}{}
\makeatletter
\relax
\def\mn@urlcharsother{\let\do\@makeother \do\$\do\&\do\#\do\^\do\_\do\%\do\~}
\def\mn@doi{\begingroup\mn@urlcharsother \@ifnextchar [ {\mn@doi@}
  {\mn@doi@[]}}
\def\mn@doi@[#1]#2{\def\@tempa{#1}\ifx\@tempa\@empty \href
  {http://dx.doi.org/#2} {doi:#2}\else \href {http://dx.doi.org/#2} {#1}\fi
  \endgroup}
\def\mn@eprint#1#2{\mn@eprint@#1:#2::\@nil}
\def\mn@eprint@arXiv#1{\href {http://arxiv.org/abs/#1} {{\tt arXiv:#1}}}
\def\mn@eprint@dblp#1{\href {http://dblp.uni-trier.de/rec/bibtex/#1.xml}
  {dblp:#1}}
\def\mn@eprint@#1:#2:#3:#4\@nil{\def\@tempa {#1}\def\@tempb {#2}\def\@tempc
  {#3}\ifx \@tempc \@empty \let \@tempc \@tempb \let \@tempb \@tempa \fi \ifx
  \@tempb \@empty \def\@tempb {arXiv}\fi \@ifundefined
  {mn@eprint@\@tempb}{\@tempb:\@tempc}{\expandafter \expandafter \csname
  mn@eprint@\@tempb\endcsname \expandafter{\@tempc}}}

\bibitem[\protect\citeauthoryear{Alexander, Clarke  \& Pringle}{Alexander
  et~al.}{2006}]{alexander2006photoevaporation}
Alexander R.,  Clarke C.,   Pringle J.,  2006, MNRAS, 369, 229

\bibitem[\protect\citeauthoryear{{Alexander}, {Pascucci}, {Andrews}, {Armitage}
   \& {Cieza}}{{Alexander} et~al.}{2014}]{alexander2013dispersal}
{Alexander} R.,  {Pascucci} I.,  {Andrews} S.,  {Armitage} P.,   {Cieza} L.,
  2014, in {Beuther} H.,  {Klessen} R.~S.,  {Dullemond} C.~P.,   {Henning} T.,
  eds, Protostars and Planets VI. pp 475--496

\bibitem[\protect\citeauthoryear{Andrews \& Williams}{Andrews \&
  Williams}{2005}]{andrews2005circumstellar}
Andrews S.~M.,  Williams J.~P.,  2005, ApJ, 631, 1134

\bibitem[\protect\citeauthoryear{Armitage}{Armitage}{2011}]{armitage2011dynamics}
Armitage P.~J.,  2011, Annual Review of Astronomy and Astroph., 49, 195

\bibitem[\protect\citeauthoryear{Bai}{Bai}{2013}]{bai2013windII}
Bai X.-N.,  2013, ApJ, 772, 96

\bibitem[\protect\citeauthoryear{Bai}{Bai}{2014}]{bai2014hall}
Bai X.-N.,  2014, ApJ, 791, 137

\bibitem[\protect\citeauthoryear{Bai}{Bai}{2017}]{bai2017global}
Bai X.-N.,  2017, ApJ, 845, 75

\bibitem[\protect\citeauthoryear{Bai \& Stone}{Bai \&
  Stone}{2013}]{bai2013windI}
Bai X.-N.,  Stone J.~M.,  2013, ApJ, 769, 76

\bibitem[\protect\citeauthoryear{Bai \& Stone}{Bai \&
  Stone}{2017}]{bai2017hall}
Bai X.-N.,  Stone J.~M.,  2017, ApJ, 836, 46

\bibitem[\protect\citeauthoryear{{Bai}, {Ye}, {Goodman}  \& {Yuan}}{{Bai}
  et~al.}{2016}]{Bai2016Thermal}
{Bai} X.-N.,  {Ye} J.,  {Goodman} J.,   {Yuan} F.,  2016, \apj, 818, 152

\bibitem[\protect\citeauthoryear{Balbus \& Hawley}{Balbus \&
  Hawley}{1991}]{balbus1991powerful}
Balbus S.~A.,  Hawley J.~F.,  1991, ApJ, 376, 214

\bibitem[\protect\citeauthoryear{B{\'e}thune, Lesur  \& Ferreira}{B{\'e}thune
  et~al.}{2017}]{bethune2017global}
B{\'e}thune W.,  Lesur G.,   Ferreira J.,  2017, A\&A, 600, A75

\bibitem[\protect\citeauthoryear{Blandford \& Payne}{Blandford \&
  Payne}{1982}]{blandford1982hydromagnetic}
Blandford R.~D.,  Payne D.,  1982, MNRAS, 199, 883

\bibitem[\protect\citeauthoryear{{Desch}}{{Desch}}{2004}]{2004ApJ...608..509D}
{Desch} S.~J.,  2004, \apj, 608, 509

\bibitem[\protect\citeauthoryear{Ercolano \& Pascucci}{Ercolano \&
  Pascucci}{2017}]{ercolano2017dispersal}
Ercolano B.,  Pascucci I.,  2017, Royal Society Open Science, 4, 170114

\bibitem[\protect\citeauthoryear{Ercolano \& Picogna}{Ercolano \&
  Picogna}{2022}]{ercolano2022modelling}
Ercolano B.,  Picogna G.,  2022, The European Phys. Journal Plus, 137, 1357

\bibitem[\protect\citeauthoryear{Ercolano, Young, Drake  \& Raymond}{Ercolano
  et~al.}{2008a}]{ercolano2008x}
Ercolano B.,  Young P.~R.,  Drake J.~J.,   Raymond J.~C.,  2008a, ApJ
  Supplement Series, 175, 534

\bibitem[\protect\citeauthoryear{{Ercolano}, {Drake}, {Raymond}  \&
  {Clarke}}{{Ercolano} et~al.}{2008b}]{2008ApJ...688..398E}
{Ercolano} B.,  {Drake} J.~J.,  {Raymond} J.~C.,   {Clarke} C.~C.,  2008b,
  \apj, 688, 398

\bibitem[\protect\citeauthoryear{{Ercolano}, {Clarke}  \& {Drake}}{{Ercolano}
  et~al.}{2009}]{2009ApJ...699.1639E}
{Ercolano} B.,  {Clarke} C.~J.,   {Drake} J.~J.,  2009, \apj, 699, 1639

\bibitem[\protect\citeauthoryear{Ercolano, Picogna, Monsch, Drake  \&
  Preibisch}{Ercolano et~al.}{2021}]{ercolano2021dispersal}
Ercolano B.,  Picogna G.,  Monsch K.,  Drake J.~J.,   Preibisch T.,  2021,
  MNRAS, 508, 1675

\bibitem[\protect\citeauthoryear{Fleming \& Stone}{Fleming \&
  Stone}{2003}]{fleming2003local}
Fleming T.,  Stone J.~M.,  2003, ApJ, 585, 908

\bibitem[\protect\citeauthoryear{Gammie}{Gammie}{1996}]{gammie1996layered}
Gammie C.~F.,  1996, ApJ, 457, 355

\bibitem[\protect\citeauthoryear{Gressel, Turner, Nelson  \& McNally}{Gressel
  et~al.}{2015}]{gressel2015global}
Gressel O.,  Turner N.~J.,  Nelson R.~P.,   McNally C.~P.,  2015, ApJ, 801, 84

\bibitem[\protect\citeauthoryear{Gressel, Ramsey, Brinch, Nelson, Turner  \&
  Bruderer}{Gressel et~al.}{2020}]{gressel2020global}
Gressel O.,  Ramsey J.~P.,  Brinch C.,  Nelson R.~P.,  Turner N.~J.,   Bruderer
  S.,  2020, ApJ, 896, 126

\bibitem[\protect\citeauthoryear{Harten, Lax  \& Leer}{Harten
  et~al.}{1983}]{harten1983upstream}
Harten A.,  Lax P.~D.,   Leer B.~v.,  1983, SIAM review, 25, 35

\bibitem[\protect\citeauthoryear{Helled et~al.,}{Helled
  et~al.}{2013}]{helled2013giant}
Helled R.,  et~al., 2013, arXiv preprint arXiv:1311.1142

\bibitem[\protect\citeauthoryear{Hollenbach, Yorke  \& Johnstone}{Hollenbach
  et~al.}{2000}]{hollenbach2000disk}
Hollenbach D.~J.,  Yorke H.~W.,   Johnstone D.,  2000, Protostars and planets
  IV, 401, 12

\bibitem[\protect\citeauthoryear{{Konigl} \& {Pudritz}}{{Konigl} \&
  {Pudritz}}{2000}]{2000prpl.conf..759K}
{Konigl} A.,  {Pudritz} R.~E.,  2000, in {Mannings} V.,  {Boss} A.~P.,
  {Russell} S.~S.,  eds, Protostars and Planets IV. p.~759

\bibitem[\protect\citeauthoryear{{Krapp}, {Gressel}, {Ben{\'{\i}}tez-Llambay},
  {Downes}, {Mohandas}  \& {Pessah}}{{Krapp}
  et~al.}{2018}]{2018ApJ...865..105K}
{Krapp} L.,  {Gressel} O.,  {Ben{\'{\i}}tez-Llambay} P.,  {Downes} T.~P.,
  {Mohandas} G.,   {Pessah} M.~E.,  2018, \apj, 865, 105

\bibitem[\protect\citeauthoryear{{Kunz}}{{Kunz}}{2008}]{2008MNRAS.385.1494K}
{Kunz} M.~W.,  2008, \mnras, 385, 1494

\bibitem[\protect\citeauthoryear{{Latter}, {Fromang}  \& {Gressel}}{{Latter}
  et~al.}{2010}]{2010MNRAS.406..848L}
{Latter} H.~N.,  {Fromang} S.,   {Gressel} O.,  2010, \mnras, 406, 848

\bibitem[\protect\citeauthoryear{Lesur, Kunz  \& Fromang}{Lesur
  et~al.}{2014a}]{lesur2014thanatology}
Lesur G.,  Kunz M.~W.,   Fromang S.,  2014a, A\&A, 566, A56

\bibitem[\protect\citeauthoryear{{Lesur}, {Kunz}  \& {Fromang}}{{Lesur}
  et~al.}{2014b}]{Lesur2014}
{Lesur} G.,  {Kunz} M.~W.,   {Fromang} S.,  2014b, \aap, 566, A56

\bibitem[\protect\citeauthoryear{Lynden-Bell}{Lynden-Bell}{1996}]{lynden1996magnetic}
Lynden-Bell D.,  1996, MNRAS, 279, 389

\bibitem[\protect\citeauthoryear{{Manara}, {Ansdell}, {Rosotti}, {Hughes},
  {Armitage}, {Lodato}  \& {Williams}}{{Manara}
  et~al.}{2023}]{2023ASPC..534..539M}
{Manara} C.~F.,  {Ansdell} M.,  {Rosotti} G.~P.,  {Hughes} A.~M.,  {Armitage}
  P.~J.,  {Lodato} G.,   {Williams} J.~P.,  2023, in {Inutsuka} S.,  {Aikawa}
  Y.,  {Muto} T.,  {Tomida} K.,   {Tamura} M.,  eds,  Astronomical Society of
  the Pacific Conference Series Vol. 534, Protostars and Planets VII. p.~539

\bibitem[\protect\citeauthoryear{Martel \& Lesur}{Martel \&
  Lesur}{2022}]{martel2022magnetised}
Martel {\'E}.,  Lesur G.,  2022, A\&A, 667, A17

\bibitem[\protect\citeauthoryear{{Nelson}, {Gressel}  \& {Umurhan}}{{Nelson}
  et~al.}{2013}]{2013MNRAS.435.2610N}
{Nelson} R.~P.,  {Gressel} O.,   {Umurhan} O.~M.,  2013, \mnras, 435, 2610

\bibitem[\protect\citeauthoryear{Owen, Ercolano, Clarke  \& Alexander}{Owen
  et~al.}{2010}]{owen2010radiation}
Owen J.,  Ercolano B.,  Clarke C.,   Alexander R.,  2010, MNRAS, 401, 1415

\bibitem[\protect\citeauthoryear{{Pascucci}, {Cabrit}, {Edwards}, {Gorti},
  {Gressel}  \& {Suzuki}}{{Pascucci} et~al.}{2023}]{2023ASPC..534..567P}
{Pascucci} I.,  {Cabrit} S.,  {Edwards} S.,  {Gorti} U.,  {Gressel} O.,
  {Suzuki} T.~K.,  2023, in {Inutsuka} S.,  {Aikawa} Y.,  {Muto} T.,  {Tomida}
  K.,   {Tamura} M.,  eds,  Astronomical Society of the Pacific Conference
  Series Vol. 534, Protostars and Planets VII. p.~567

\bibitem[\protect\citeauthoryear{Perez-Becker \& Chiang}{Perez-Becker \&
  Chiang}{2011}]{perez2011surface}
Perez-Becker D.,  Chiang E.,  2011, ApJ, 735, 8

\bibitem[\protect\citeauthoryear{{Picogna}, {Ercolano}, {Owen}  \&
  {Weber}}{{Picogna} et~al.}{2019}]{picogna2019dispersal}
{Picogna} G.,  {Ercolano} B.,  {Owen} J.~E.,   {Weber} M.~L.,  2019, \mnras,
  487, 691

\bibitem[\protect\citeauthoryear{{Ribas}, {Mer{\'\i}n}, {Bouy}  \&
  {Maud}}{{Ribas} et~al.}{2014}]{Ribas2014}
{Ribas} {\'A}.,  {Mer{\'\i}n} B.,  {Bouy} H.,   {Maud} L.~T.,  2014, \mn@doi
  [\aap] {10.1051/0004-6361/201322597}, \href
  {https://ui.adsabs.harvard.edu/abs/2014A&A...561A..54R} {561, A54}

\bibitem[\protect\citeauthoryear{Rodenkirch, Klahr, Fendt  \&
  Dullemond}{Rodenkirch et~al.}{2020}]{rodenkirch2020global}
Rodenkirch P.~J.,  Klahr H.,  Fendt C.,   Dullemond C.~P.,  2020, A\&A, 633,
  A21

\bibitem[\protect\citeauthoryear{Stone, Gammie, Balbus  \& Hawley}{Stone
  et~al.}{2000}]{stone2000transport}
Stone J.~M.,  Gammie C.~F.,  Balbus S.~A.,   Hawley J.~F.,  2000, Protostars
  and Planets IV, 589

\bibitem[\protect\citeauthoryear{Wang, Bai  \& Goodman}{Wang
  et~al.}{2019}]{wang2019global}
Wang L.,  Bai X.-N.,   Goodman J.,  2019, ApJ, 874, 90

\bibitem[\protect\citeauthoryear{{Wardle} \& {Ng}}{{Wardle} \&
  {Ng}}{1999}]{1999MNRAS.303..239W}
{Wardle} M.,  {Ng} C.,  1999, \mnras, 303, 239

\bibitem[\protect\citeauthoryear{{Wardle} \& {Salmeron}}{{Wardle} \&
  {Salmeron}}{2012}]{wardle2012hall}
{Wardle} M.,  {Salmeron} R.,  2012, \mnras, 422, 2737

\bibitem[\protect\citeauthoryear{Ziegler}{Ziegler}{2004}]{ziegler2004adi}
Ziegler U.,  2004, Computer Phys. Comm., 157, 207

\bibitem[\protect\citeauthoryear{Ziegler}{Ziegler}{2016}]{ziegler2016chemical}
Ziegler U.,  2016, A\&A, 586, A82

\makeatother
\end{thebibliography}


\bsp	
\label{lastpage}

\end{document}